\documentclass[12pt]{article}
\usepackage{color}
\usepackage{latexsym}
\usepackage{epsfig,amssymb,euscript, mathrsfs}
\usepackage{amsmath}
\newsavebox{\ns}
\newsavebox{\dbrane}
\newsavebox{\dbshort}

\usepackage{comment}
\usepackage{amsmath}
\usepackage{hyperref}

\newcommand\R{\mathbb{R}}

\newcommand\diff{\mathrm{d}}

\newcommand{\ti}{\widetilde}
\newcommand{\half}{\frac{1}{2}}
\newcommand{\trho}{\rho}

\newcommand{\sq}{v}

\newcommand{\bR}{\ensuremath{\mathbb{R}}}

\newcommand{\bZ}{\ensuremath{\mathbb{Z}}}

\newcommand{\scL}{\ensuremath{\mathcal{L}}}

\newcommand{\scO}{\ensuremath{\mathcal{O}}}

\newcommand{\scW}{\ensuremath{\mathcal{W}}}

\newcommand{\fb}{\mathfrak b }

\newcommand{\lp}{\left(}
\newcommand{\rp}{\right)}

\newlength{\sswidth}

\numberwithin{equation}{section}       

\newcommand{\bea}{\begin{eqnarray}}
\newcommand{\eea}{\end{eqnarray}}
\newcommand{\be}{\begin{equation}}
\newcommand{\ee}{\end{equation}}

\newcommand{\ba}{\begin{equation}\begin{array}}
\newcommand{\ea}{\end{array}\end{equation}}

\newcommand{\nn}{\nonumber}

\def \bfone{\relax{\textrm 1\kern-.35em 1}}

\newcommand{\timp}{Y}

\numberwithin{equation}{section}       

\begin{document}

\begin{titlepage}
\begin{flushright}
KCL-MTH-15-02 
\end{flushright}
\bigskip
\def\thefootnote{\fnsymbol{footnote}}

\begin{center}
\vskip -5mm

{\LARGE
{\bf The Casimir Energy in Curved Space \\ \vspace{0.3 em} and its Supersymmetric Counterpart}
}
\end{center}

\vskip 5mm 

\begin{center}
{\large
Benjamin Assel,$^1$
Davide Cassani,$^2$
Lorenzo Di Pietro,$^3$ \\
\vskip 5pt
Zohar Komargodski,$^3$
Jakob Lorenzen,$^1$ and
Dario Martelli$^1$}

\end{center}

\renewcommand{\thefootnote}{\arabic{footnote}}

\begin{center}
$^1$ {\it Department of Mathematics, King's College London, \\ [1mm]
The Strand, London WC2R 2LS,  United Kingdom\\ [2mm]}
$^2$ {\it Sorbonne Universit\'es UPMC Paris 06,\\ [1mm] 
UMR 7589, LPTHE, F-75005, Paris, France\\ [1mm]
and\\ [1mm]
CNRS, UMR 7589, LPTHE, F-75005, Paris, France\\ [2mm]}
$^3$ {\it Department of Particle Physics and Astrophysics,\\ [1mm] Weizmann Institute of Science,
Rehovot, 76100, Israel\\}

\end{center}

\vskip 1cm

\noindent
\begin{center} {\bf Abstract} \end{center}

We study $d$-dimensional Conformal Field Theories (CFTs) on the cylinder, $S^{d-1}\times \R$, and its deformations. In $d=2$ the Casimir energy (\emph{i.e.}\ the vacuum energy) is universal and is related to the central charge $c$. In $d=4$ the vacuum energy depends on the regularization scheme and has no intrinsic value. We show that this property extends to infinitesimally deformed cylinders and support this conclusion with a holographic check. 
However, for ${\cal N}=1$ supersymmetric CFTs, a natural analog of the Casimir energy turns out to be {\it scheme independent} and thus intrinsic. We give two proofs of this result. We compute the Casimir energy for such theories by reducing to a problem in supersymmetric quantum mechanics. For the round cylinder the vacuum energy is proportional to $a+3c$. We also compute the dependence of the Casimir energy on the squashing parameter of the cylinder. Finally, we revisit the problem of supersymmetric regularization of the path integral on Hopf surfaces.

\vfill

\end{titlepage}

\pagestyle{plain}
\setcounter{page}{1}
\newcounter{bean}
\baselineskip18pt
\tableofcontents


\section{Introduction and summary}

The quantum anomalies appearing in the trace of the energy-momentum tensor encode  universal information about  Conformal Field Theories (CFTs) in even space-time dimensions. In $d=2$  
the conformal anomaly $c$ characterizes 
conformal field theories~\cite{Cardy:2010fa}, and  constrains the renormalization group (RG)  flows between  them \cite{Zamolodchikov:1986gt}. In four-dimensional CFTs, there are two trace anomaly coefficients, $c$ and $a$.
They appear in many applications. In particular, Cardy conjectured \cite{Cardy:1988cwa} that there exists a decreasing function along RG flows, reducing to $a$ at the fixed points~\cite{Jack:1990eb,Komargodski:2011vj,Komargodski:2011xv}. 
 
 Here we will be interested in the following question. Given a conformal field theory in $\mathbb{R}^{d}$, $d=2n$, we can use a Weyl transformation and study the theory on $S^{d-1}\times \mathbb{R}$. This is often referred to as ``radial quantization.''
 Denoting the noncompact coordinate by $\tau$, we can ask about the energy of the ground state $E_0$, defined as 
\be
E_0 \ =\ \int_{S^{d-1}} \diff^{d-1}x \sqrt g \, \langle T_{\tau\tau} \rangle~.
\label{cas}
\ee
The expectation value of the energy-momentum tensor is taken in the ground state of the CFT on the cylinder. We refer to $E_0$ as the Casimir energy.  

It is crucial for the analysis below to understand whether $E_0$ is well defined, namely,  scheme independent.  In $d=2$ the only dimensionless counterterm is 
 \be
\int \diff^2x \sqrt{g} R~,
\ee
where $R$ denotes the Ricci scalar. This vanishes on the cylinder and thus does not shift the vacuum energy.\footnote{More precisely, since there is the cosmological constant counterterm $\Lambda^2_{UV}\int \diff^2x \sqrt{g}$, the vacuum energy would generally have power divergent pieces $E_0\sim \Lambda_{UV}^2 r_1$ (with $r_1$ the radius of the $S^1$ and $\Lambda_{UV}$ the UV cut-off) which are non-universal. So when we discuss the scheme-independence of the vacuum energy, we always have in mind the piece that remains finite when the cut-off is removed. In $d=2n$ dimensions, only counterterms with $d$-derivatives of the metric are thus relevant to us.\label{foot:counterterms}}
In $d=4$ there are several dimensionless counterterms. A basis is given by the Euler density $E_{(4)}$, the square of the Weyl tensor $W^2$, the Pontryagin density ${\rm Tr} (R\wedge R)$, and $R^2$. Of those four dimensionless counterterms only $\int \diff^4 x \sqrt{g} R^2$ does not vanish on $S^3\times \mathbb{R}$.  We could thus add this counterterm to the action with some scheme-dependent coefficient $b$
\be
\delta S \ =\ - {b\over 12(4\pi)^2} \int \diff^4 x \sqrt{g} R^2~.
\label{localterm}
\ee
The curvature of the three-sphere of radius $r_3$ is $R={6\over r_3^2}$ in our conventions. As a consequence, integrating the counterterm above we get 
$\delta S= - {3 b\over 8r_3}\int \diff\tau$. Interpreting the coefficient of $\int \diff\tau$ as the ground state energy we see that $E_0$ is ambiguous. An equivalent way to understand this ambiguity is to note that~\eqref{localterm} leads to a scheme-dependent term in the trace of the energy-momentum tensor 
\bea
\langle T_\mu^\mu \rangle & = & \frac{1}{(4\pi)^2}\left(a E_{(4)} - c W^2   + b \Box R \right) ~.
\label{tracet}
\eea
Since $\langle T_\mu^\mu \rangle$ is modified, this affects the other components of $\langle T_{\mu
\nu} \rangle$ and in particular the vacuum energy.
By contrast, in $d=2$ no dimension 2 term can be added to the right-hand side of $\langle T_\mu^\mu \rangle =-\frac{c}{24\pi}R$. 
 
We see that,  without additional assumptions, the ground state energy on $S^3\times \R$ is not an intrinsic property of the CFT. It depends on the regularization scheme, \emph{i.e.}\ the precise way in which the CFT is defined (different definitions lead to different values of $b$). On the other hand, the Casimir energy on $S^1\times \R$ is an intrinsic observable that does not depend on the ultraviolet completion. 
 
For completeness, let us quote the results for $E_0$ in $d=2$ and in $d=4$. In $d=4$ below we use the general scheme with some $b$. We present a self-contained derivation of these results in Appendix A. 
The Casimir energy on $S^1\times \R$ is
 \be\label{C2d}
  E_0 \ = \  - \frac{c}{12 r_1}~ ,
 \ee
 where $r_1$ is the radius of the circle. The Casimir energy on $S^3\times \R$ is 
\be
 E_0 \ = \ \frac{3}{4r_3} \left(a-\frac{b}{2}\right) ~,
 \label{nobox}
 \ee
where $r_3$ is the radius of the three-sphere. 
In fact, as we show in Appendix A, the result in $d=4$ remains ambiguous even if one allows for an infinitesimal deformation of the three-sphere. By this we mean that
the linear response to an infinitesimal deformation is still proportional to $a-{b\over 2}$.  We also provide a holographic test of this claim, by checking it against the linearized holographic Casimir energy of the supergravity solution of~\cite{Cassani:2014zwa}. 
  
The three main points of the present paper are 
\begin{itemize}
\item There is a natural generalization of the notion of Casimir energy for ${\cal N}=1$ supersymmetric field theories on $S^3\times \R$. 
\item It turns out that this Casimir energy is {\it physical}, \emph{i.e.}\ scheme independent. We will give two proofs of this assertion and then we will evaluate the Casimir energy exactly. One proof is based on~\cite{Assel:2014tba} and the other is based on ideas related to Chern-Simons terms in quantum mechanics. The Casimir energy comes out proportional to the linear combination $a+3c$ of the conformal anomalies. 
\item When one computes supersymmetric partition functions on manifolds with $S^3\times S^1$ topology, the answer is given in terms of a product of the Casimir factor and the usual supersymmetric index. This leads to consistent results in various limits of the partition function. 
\end{itemize}  
  
Let us now briefly explain why it might be useful to understand the Casimir factor for four-dimensional theories. Recall that in $d=2$, in order for the torus partition function to be modular invariant, it is necessary to include the Casimir prefactor $e^{\frac{c\beta}{12r_1}}$ (where $\beta$ is the length of the thermal circle). The Casimir prefactor in four dimensions might be similarly important to manifest various relations between different path integrals such as those analogous to~\cite{Pasquetti:2011fj} and {\it e.g.}~\cite{Yoshida:2014qwa}. The Casimir prefactor might be also relevant for a better understanding of the role of the modular-like transformations in four dimensions discussed in~\cite{Spiridonov:2012ww,Razamat:2012uv} (and see references therein).

So let us begin by discussing how we put supersymmetric ${\cal N}=1$ theories on $S^3\times \R$. 
For theories with an $R$-symmetry (such as any SCFT) one can follow the procedure of~\cite{Festuccia:2011ws} whereby one couples the theory to background new minimal supergravity. New minimal supergravity~\cite{Sohnius:1981tp} contains the bosonic fields $g_{\mu\nu},A_\mu, B_{\mu\nu}$, where $A_\mu$ is the $R$-gauge field and $B_{\mu\nu}$ is a two-form.\footnote{One is allowed to discuss the $R$-symmetry in curved space without any significant modifications because its gravitational anomaly vanishes under very general assumptions \cite{Cassani:2013dba}.} The procedure works both in Lorentzian and in Euclidean signature; here we refer to the Euclidean case for definiteness.

It turns out that for every complex manifold with a Hermitian metric one can find appropriate $A_\mu$ and $B_{\mu\nu}$ such that one preserves at least one rigid supercharge \cite{Klare:2012gn,Dumitrescu:2012ha}. This can be viewed as a generalization of the notion of twisting~\cite{Closset:2014uda}. Furthermore, the partition function is independent of the Hermitian metric; it only depends on the complex structure of the underlying complex manifold~\cite{Closset:2013vra}.
A particularly interesting class of complex manifolds are primary Hopf surfaces, which are topologically equivalent to $S^3\times S^1$. In this case, the relevant branch of the moduli space of complex structures is two-complex dimensional, parameterized by complex numbers $p,q$. 

For the special case of the round metric on $S^3\times S^1$, the complex structure parameters are given in terms of the $S^1$-length $\beta$ and the $S^3$-radius $r_3$ as $p= q = e^{- \frac{\beta}{r_3} }$. This geometry preserves in fact four supercharges and in addition to the round metric we need to activate the background fields $A_\tau=V_\tau=\tfrac{i}{r_3}$, where $V_\mu={1\over 4} \epsilon_{\mu\nu\rho\sigma}\nabla^{\nu}B^{\rho\sigma}$. The two-form $B_{\mu\nu}$ couples to a redundant operator in conformal field theories and is not so important. However, the flat gauge field $A$ along the Euclidean time direction implies that the Hamiltonian, $H_{\rm susy}$, is shifted with respect to what we would get from radial quantization according to 
\be\label{SUSYH}
H_\mathrm{susy} \ =\ \Delta - \frac{1}{2r_3} R~.
\ee 
Above $\Delta$ is the time translation operator that is obtained by mapping the dilatation operator in flat space to the cylinder (equivalently, $\Delta$ is the canonical Hamiltonian for a theory on the cylinder with $A_\tau=\frac{3i}{2r_3},\, V_\tau=\tfrac{i}{r_3}$). The Hamiltonian~\eqref{SUSYH} commutes with the four supercharges on the round $S^3\times S^1$ and so the Hilbert space is organized in representations of the appropriate superalgebra. 

 The Casimir energy, $E_{\rm susy}$, is thus naturally defined from the large $\beta$ limit of the partition function~\cite{Assel:2014paa} 
 \bea
Z_{ S^3\times S^1_\beta}^\mathrm{susy}  & \sim & e^{- \beta E_\mathrm{susy}} \qquad\quad  \mathrm{for}  \qquad\quad \beta \to \infty ~.
\label{defsusycasso}
\eea
In other words, $E_\mathrm{susy}$ is the lowest eigenvalue of the Hamiltonian~\eqref{SUSYH}. See also Appendix \ref{app:EvsH}. 

This definition can be extended to manifolds  $M_3\times S^1$, with $M_3$ a Seifert manifold (a Seifert manifold is, loosely speaking, a circle bundle over a Riemann surface). For primary Hopf surfaces, we will deform the  three-sphere with some squashing parameter $\mathfrak b$, 
related to the underlying complex structure parameters $p,q$,  and study $E_{\mathrm{susy}}(\mathfrak b)$. Our final result for the round sphere ($\fb=1$) with radius $r_3$ is\footnote{The formula also applies to non-conformal theories by replacing $a$ and $c$ with the appropriate linear combinations of traces over the fermion $R$-charges ${\rm Tr} R$ and ${\rm Tr} R^3$ (see~\eqref{AnomCoeff}).}
\bea
E_\mathrm{susy} & = & \frac{4}{27 r_3}( a + 3  c) ~,
\label{secondesusy}
\eea
in agreement with~\cite{Kim:2012ava}. The partition function is therefore given by 
\bea \label{pref}Z_{ S^3\times S^1_\beta}^\mathrm{susy} \ = \ e^{-\frac{4\beta}{27 r_3}( a + 3  c)} {\cal I}_{S^3\times S^1_\beta}~, \eea with   $ {\cal I}_{S^3\times S^1_\beta} $ being the usual supersymmetric index~\cite{Romelsberger:2005eg, Kinney:2005ej, Romelsberger:2007ec, Dolan:2008qi}.
The results of~\cite{DiPietro:2014bca, Ardehali:2014zba, Ardehali:2014esa, Ardehali:2015hya} are consistent with~\eqref{pref}.

Since the Casimir energy is not well defined without supersymmetry, one has to develop a supersymmetric method to regularize the a priori divergent Casimir energy. As we will explain in the following subsection, the main idea is that the expectation value of~\eqref{SUSYH} can be related to the expectation value of the $R$-charge in the vacuum. The latter can be thought of as a Chern-Simons term in quantum mechanics and this leads to a determination of the ordering ambiguities (see also~\cite{Lorenzen:2014pna}).

Let us mention that recovering  (\ref{secondesusy}) from a holographic computation remains an open problem. Recall that for four-dimensional SCFTs admitting a weakly coupled gravity dual, at leading order in the large $N$ limit, one has $a=c$ and thus we predict
$E_\mathrm{susy}  =  \frac{16}{27 r_3}a \sim  {\cal O}(N^2)$. Therefore this should be reproduced by the on-shell action of an appropriate supergravity solution. 
It was noted in~\cite{Cassani:2014zwa} that an obvious
candidate for such a solution is AdS$_5$, with the addition of an appropriate flat background gauge field; however, its renormalised on-shell action is the same as that of pure 
AdS$_5$, which agrees with (\ref{nobox}) (with $b=0$) \cite{Herzog:2013ed}, and does not agree with (\ref{secondesusy}). 
It is clearly worthwhile to revisit this problem.

In the next subsection we briefly summarize the derivation of~\eqref{secondesusy}. Section~2 contains the detailed derivation of~\eqref{secondesusy} and Section~\ref{DeformedS3} contains the generalization to squashed cylinders.  
Appendix A contains a study (which is mostly a review) of the non-supersymmetric Casimir energy. In Appendix~\ref{app:EvsH} we include  some details  about the identification of $E_{\rm susy}$ with the VEV of the Hamiltonian. Finally, in Appendix~\ref{app:1LoopDet} we reconsider the full partition function on $S^3\times S^1$. We revisit the regularization of one-loop determinants and determine the relation of the partition function and the supersymmetric index for all $p,q$, a study initiated in~\cite{Assel:2014paa} (see also~\cite{Closset:2013sxa}), thereby generalizing~\eqref{pref}. The correct regularization of the one-loop determinants  leads to results consistent with the Cardy-like formula~\cite{DiPietro:2014bca} (see also \cite{Ardehali:2014zba}), namely
\bea
Z_{S^3\times S^1_\beta}^\mathrm{susy}  & \sim &    \exp \left(-\frac{16\pi^2 r_3}{3\beta} (a-c) \right)      \qquad\quad  \mathrm{for}   \qquad\quad\beta \to 0 ~,
\eea
as advertised in \cite{DiPietro:2014bca,Lorenzen:2014pna}.


\subsection{Outline of the derivation}

Let us now summarize the main points in the derivation of the supersymmetric Casimir energy for SCFTs.

In section \ref{sec:SusyVacE} we consider an $\mathcal N=1$ theory with an $R$-symmetry on the conformally flat background $S^3 \times \R$, with round metric on $S^3$. We work in Lorentzian signature and denote the real time $t = i \tau$. Preserving supersymmetry requires to turn on background fields of the new minimal supergravity multiplet $A_t = V_t = \tfrac{1}{r_3}$. Focusing on two supercharges of opposite $R$-charge (out of the four preserved by the background), the supersymmetry algebra on $S^3\times\R$ takes the form
\be
\frac{1}{2} \{Q,Q^\dagger \} \ = \  H_\mathrm{susy} -  \frac{1}{r_3}( R + 2J_3)    ~ , \qquad  {[}H_\mathrm{susy},Q]  \ = \ {[}R +2 J_3,Q] \ = \ 0\ .
\ee

Here $R$ is the generator of $U(1)_R$ symmetry while $J_3$ is a Cartan generator of the isometries of the three-sphere. The Casimir energy is given by the VEV of the Hamiltonian appearing in the algebra, $E_\mathrm{susy} = \langle H_\mathrm{susy} \rangle $. Here, the expectation value is evaluated in the vacuum of the theory coupled to the various background fields. Since the supercharge is time-independent, we can then Wick rotate and compactify the time direction on a circle of length $\beta$, and $E_{\rm susy}$ gives the dominant contribution to the supersymmetric partition function for $\beta \to \infty$ as in \eqref{defsusycasso}. 

Because of the separation of scales between the radius of the three-sphere and the radius of the circle, it is natural to study the reduction of the theory on the three-sphere. The result of this reduction is a \emph{supersymmetric quantum mechanics} with infinitely many degrees of freedom. We are thus led to consider a $1d$ system with Hamiltonian $H_\mathrm{susy}$, a global symmetry with charge $\Sigma$ and a supersymmetry algebra of the form 
\bea\label{SUSYQMintro}
\{Q,Q^\dagger \} & = &  2 \Big(H_\mathrm{susy} - \frac{1}{r_3} \Sigma \Big)    ~ , \qquad \quad {[}H_\mathrm{susy}, Q] \ =  \  {[}\Sigma, Q]  \ = \ 0  ~.
\eea
Unbroken supersymmetry implies that $r_3 \langle H_\mathrm{susy} \rangle = \langle \Sigma \rangle $. 
When we reduce to $1d$,  the effective action that computes $\langle \Sigma \rangle$ is given by
\be\label{QMcounterintro}
\mathcal W[A^{\Sigma}_t]\ =\ \langle \Sigma\rangle \int \diff t  \, A^{\Sigma}_t \ ,
\ee
where $A^\Sigma$ is the background gauge field for the symmetry generated by $\Sigma$. 
This is a $1d$ Chern-Simons (CS) term associated to the VEV of $\Sigma$, {\it i.e.}\ the charge of the vacuum.

 In quantum mechanics \eqref{QMcounterintro} can in principle be completed to a supersymmetric counterterm, involving the $1d$ metric $g_{tt}$ as well as other background supergravity fields.  Such a counterterm would reflect the fact that $\langle H_\mathrm{susy} \rangle$ and $\langle \Sigma \rangle$ can be shifted by an arbitrary amount, preserving the relation $r_3 \langle H_\mathrm{susy} \rangle = \langle \Sigma \rangle $ imposed by the superalgebra \eqref{SUSYQMintro}. By contrast, if the quantum mechanical model arises from a local higher-dimensional model, the allowed counterterms must descend from local higher-dimensional counterterms. It easy to see that the quantum-mechanical CS term~\eqref{QMcounterintro} cannot descend from a higher-dimensional counterterm and thus it is scheme independent. As a consequence, since the vacuum energy is fixed by supersymmetry to be the same  as the $1d$ CS term, the vacuum energy is physical. Furthermore,
since the CS term cannot depend on continuous coupling constants, the vacuum energy is also independent of continuous coupling constants. If we further assume the existence of a weakly coupled point, we can reliably compute the Casimir energy using a free field theory.

We can then consider a free chiral multiplet. Supersymmetry implies that upon reduction on the $S^3$ the resulting Lagrangian and supersymmetry transformations 
must be organized in terms of $1d$ multiplets and Lagrangians. In particular, there are two types of  multiplets, that we will call ``short.'' 
These are the {\it chiral} $(\phi, \psi)$ and the {\it Fermi} $(\lambda, f)$ multiplets. A key point is that the combination of the two multiplets can form a reducible but indecomposable representation of supersymmetry. Thus they can join to form a ``long'' multiplet, with the coupling between the short components controlled by a parameter $p$. When $p=0$, the short multiplets are decoupled.
  
The scalar fields of the $1d$ chiral multiplets arise from ``Fourier modes'' of the scalar fields in the $4d$ chiral multiplet. For example, on the round sphere we can use the spherical harmonics and write $\phi =  \sum \phi_{\ell,m,n} Y_{\ell,m,n}\,.$
Here $\ell, m, n$ are the usual $SU(2)_l \times SU(2)_r$ quantum numbers. 
Similarly, one can expand the two-component fermion of the $4d$ chiral multiplet in spinorial harmonics, and the $1d$ fermions $\psi$ and $\lambda$ may be identified with linear combinations of its components. 
The deformation parameter $p$ that governs the shortening of the multiplets is related to the $4d$ quantum number as  $p^2 \ = \ (\ell - 2m )(\ell + 2 + 2m)$, 
hence the shortening condition $p=0$ is satisfied if and only if $m = \ell/2$ or $m = -\ell/2 -1$.

We will see that in the correct renormalization scheme (whose existence we established above), the Hamiltonian of a long multiplet exactly \emph{vanishes} on the vacuum, while for the short multiplets we find 
\bea
\mathrm{chiral}~~(m=\frac{\ell}{2}): & \langle H_\mathrm{chiral} \rangle   \ =\   \frac{1}{2 r_3} (\ell +r)~,   \nonumber\\
\mathrm{Fermi}~~(m=-\frac{\ell}{2}-1): &\qquad  ~\langle H_\mathrm{Fermi} \rangle  \ =\  -  \frac{1}{2 r_3} (\ell +2- r)  \ ,
\eea
where $r$ is the $R$-charge of the $4d$ chiral multiplet.
Thus, given that the long multiplets do not contribute, the expectation value of the total Hamiltonian is 
\be\label{Hchiral+Hfermi}
\langle  H_\mathrm{susy} \rangle  \ = \  \sum_\mathrm{chiral} \langle H_\mathrm{chiral}\rangle + \sum_\mathrm{Fermi} \langle H_\mathrm{Fermi} \rangle\ .
\ee
The two infinite sums can be regularized using different methods, for example using the Hurwitz zeta function, or a cut-off method, (see also \cite{Assel:2014paa} and \cite{Lorenzen:2014pna}) yielding in any case the 
result\footnote{In this paper we do not carry out the explicit analysis for the vector multiplets; the results would be consistent with our conclusions. Additionally, the combination $a + 3c$ has an interesting interpretation in terms of the anomaly polynomial~\cite{BobevBullimoreLee}.}  
\be
\langle  H_\mathrm{susy} \rangle \ =\ E_\mathrm{susy} \ =\ \frac 4 {27r_3} (a + 3c)~.
\ee

Using a similar strategy, in section 3 we compute the supersymmetric Casimir energy for an ${\cal N}=1$ theory (with an $R$-symmetry) on a class of Hopf surfaces with 
$U(1)^3$ symmetry, considered in  \cite{Assel:2014paa}. Let us describe this briefly. Again we start from the free chiral multiplet Lagrangian in four dimensions, and
reduce this to an infinite set of fields in $1d$, comprising long and short multiplets. Since the quantized Hamiltonian of long multiplets vanishes on the vacuum, we can focus on the short multiplets. The shortening conditions are expressed as first-order differential equations for the eigenstates of the Hamiltonians for chiral and Fermi multiplets, which are labeled by two non-negative integers $n_1,n_2$. This can be seen using the method of 
\cite{Alday:2013lba} or, equivalently, utilising  the variables of \cite{Closset:2013sxa}. Each chiral and Fermi multiplet contributes as
\bea
\mathrm{chiral}:~~ &\;\, \langle H_\mathrm{chiral} \rangle \!  \ =\  \frac{1}{2r_3}\left(  | \fb|n_1 + |\fb^{-1}| n_2 + \frac{r}{2}(|\fb| + |\fb^{-1}|) \right) \ ,\nonumber\\ [1mm]
\mathrm{Fermi}:~~ & \qquad \langle  H_\mathrm{Fermi}  \rangle\! \ = \ -\frac{1}{2r_3}\left( |\fb| n_1 + |\fb^{-1}|n_2 + \frac{2-r}{2}(|\fb| + |\fb^{-1}|) \right)~,
\eea
where $\fb$ is a parameter deforming the complex structure of the Hopf surface.
Again the two sums in \eqref{Hchiral+Hfermi} must be regulated separately. This can be done using the Barnes double zeta function, which is a two-parameter 
generalization of the Hurwitz zeta function. In the end, we obtain
\be\label{squashedC}
\langle H_\mathrm{susy} \rangle \ = \ \frac{2}{3r_3}\left(|\fb| + |\fb|^{-1} \right)({a}-{ c}) + \frac{2}{27 r_3} (|\fb| + |\fb|^{-1} )^3 (3\,{ c}-2\,{ a}) \ ,
\ee
which is  the expression (5.10) in  \cite{Assel:2014paa} (with the redefinition $b_1 = \frac{\beta \fb}{ 2\pi r_3}$ and $b_2 = \frac{\beta \fb^{-1}}{ 2\pi r_3 }$).


\section{Supersymmetric Casimir energy}
\label{sec:SusyVacE}

In this section we perform a manifestly supersymmetric analysis of the Casimir energy problem on $S^3 \times \mathbb R$.
Our strategy is to reduce to a one-dimensional quantum mechanical problem. Using the same idea, in section~\ref{DeformedS3} we will discuss the more general case leading to~\eqref{squashedC}. 
 

\subsection{Definition}
\label{DefCasimir}

Consider an ${\mathcal N}=1$ theory with an $R$-symmetry. The energy-momentum tensor can be embedded in the so-called $R$-multiplet (see, for instance,~\cite{Komargodski:2010rb}), hence the theory can be coupled canonically to new minimal supergravity. The bosonic part of the supergravity multiplet consists of the metric $g_{\mu\nu}$ and two auxiliary fields, the $R$-gauge field $A_\mu$ and a conserved one-form $V_\mu$, $\nabla_\mu V^\mu =0$, dual to the field strength of a two-form gauge field $B_{\mu\nu}$. 
Gravity can then be made non-dynamical by taking a rigid limit, so that one is left with an ordinary field theory on a four-manifold $M_4$. In this limit generically all the supergravity fields above are turned on, and play the role of background fields.  For particular choices of background fields and manifolds $M_4$ one can show that the theory on $M_4$ preserves some rigid supersymmetry. The condition is that there exists a nontrivial solution to the generalized Killing spinor equation~\cite{Festuccia:2011ws}
\be
\label{ksei} (\nabla_\mu -i A_\mu) \zeta_\alpha + iV_\mu \zeta_\alpha +  iV^\nu (\sigma_{\mu\nu}\zeta)_\alpha \ =\ 0~,
\ee
and its conjugate equation. 
In Lorentzian signature, these equations admit a solution if and only if the background admits a null Killing vector~\cite{Cassani:2012ri}. In Euclidean signature, solutions were classified in~\cite{Klare:2012gn,Dumitrescu:2012ha}. It turns out that a solution exists if and only if the manifold is complex, with a Hermitian metric. For some special cases (and the cases we analyze in this paper are of this type) one can naturally Wick rotate from the Euclidean to the Minkowski solution. Examples of such backgrounds are $M_4 = M_3 \times S^1$, with $M_3$ a Seifert manifold (examples include spheres and Lens spaces).

An $\mathcal N=1$ theory on a curved manifold $M_4$ presents the same short-distance behavior as in flat space. So in general there will be divergences in the computation of {\it e.g.}\ the partition function on $M_4$, which need to be renormalized. Assuming that the theory is regularized in a way which respects both supersymmetry and the $R$-symmetry, we infer that two different regularization schemes differ by some local counterterm, namely some local action constructed using the background new minimal supergravity fields, $\int \diff^4x\sqrt g\,{\cal L}(g_{\mu\nu},A_\mu,V_\mu)$. This has to respect supersymmetry and gauge invariance. Local terms of dimension four affect the finite part of the partition function, as they remain finite when the UV cut-off is removed. In \cite{Assel:2014tba} it was shown that when there exist two Killing spinors of opposite chirality, then all local counterterms of dimension four evaluate to zero. This means that such partition functions are in fact scheme-independent.\footnote{As in footnote~\ref{foot:counterterms}, one should bear in mind that there is a dimension $<$ 4 counterterm that does not vanish in general, but this contains a positive power of the cut-off. We will use such divergent counterterm in appendix~\ref{app:1LoopDet} to regularize the partition function.} 

The results of \cite{Closset:2013vra,Closset:2014uda} tell us that $Z^{\rm susy}_{M_3\times S^1_\beta}$ is independent of coupling constants and only depends on complex structure parameters of the underlying complex four-manifold $M_3\times S^1$. The findings of this work are consistent with that.

The partition function can be interpreted as usual as a trace over the Hilbert space $Z^{\rm susy}_{M_3\times S^1_\beta} = {\rm Tr}\left[(-)^F e^{-\beta H_{\rm susy}}\right]$, where $H_{\rm susy}$ generates the time evolution along $S^1$. 
Then the Casimir energy is extracted from the large circle limit as
\be
E_{\rm susy} \ = \ -\lim_{\beta\to \infty}\frac{{\rm d}}{{\rm d}\beta}\log Z^{\rm susy}_{M_3 \times S^1_\beta}\ .
\ee
The scheme-independence of the partition function implies that the vacuum energy is universal (below we will also give a new proof of this fact). This should be contrasted with the non-universality of the ordinary Casimir energy $E_0$ discussed in the introduction.


\subsection{Consequences of the supersymmetry algebra}

In this section we make a few observations based on the supersymmetry algebra. 
In Lorentzian signature, let us consider the simple case $M_4 =S^3 \times \mathbb R$, where $S^3$ is a round three-sphere of radius $r_3$ and $\mathbb R$ is the time direction parameterized by $t$ (this is related to the Euclidean time $\tau$ in the introduction by $t =i\tau$). By fixing the other background fields to $A_t = V_t = \frac{1}{r_3}$, with all other components vanishing, we can preserve four supercharges\footnote{This requires making a special choice of the so-called $\kappa$ parameter~ \cite{Lorenzen:2014pna}. For generic choices 
of $\kappa$, the background preserves only two supercharges.} for any $\mathcal N=1$ theory with an $R$-symmetry~\cite{Festuccia:2011ws}. In fact one could take any constant $A_t$ for the flat gauge field. However, only in the case that $A_t = \frac{1}{r_3}$ one gets time-independent supercharges. So we will make this choice throughout.

The superalgebra preserved by the background is~\cite{Romelsberger:2005eg} 
\bea
\label{supali} \frac 12 \{Q_\alpha,  Q^{\dagger}{}^{\beta}\}  &=&  \delta{}^{\beta}{}_{\alpha}  \Big(  H_{\rm{susy}}-\frac{1}{r_3} R   \Big) - \frac{2}{r_3} \sigma^i \, {}^{\beta}{}_{\alpha}  J_l^i~,\nn \\ [2mm]
\label{supalii}[H_{\rm{susy}},Q_\alpha]  &=& 0~, \quad [R,Q_\alpha] \ = \ - Q_\alpha \, , \quad [J_l^i , Q_\alpha] \ = \ \frac 12 \, Q_\beta \,  \sigma^i \, {}^{\beta}{}_{\alpha} ~,
\eea
where $H_{\rm susy}$ is the generator of translations along the circle, $R$ is the $R$-charge, the $J_l^i$, $i=1,2,3$, are the generators of the $SU(2)_l\subset SU(2)_l\times SU(2)_r $ isometry of the sphere and $\sigma^i$ are the Pauli matrices. The supercharges $Q_\alpha$, $\alpha =1,2$, form a doublet of $SU(2)_l$, while the $SU(2)_r$ subgroup does not appear in the superalgebra. 

A first remark is that  we assume {\it the vacuum does not break supersymmetry.} Suppose the vacuum were not supersymmetric, in which case either $Q_1$, or $Q_2$ (or both) would not annihilate the vacuum. Then, $Q|{\rm VAC}\rangle$ is a new state with the same value of $H_{\rm susy}$, but contributing with an opposite sign to the index or partition function. Therefore if supersymmetry were broken, the index on $S^3\times \mathbb R$ would not receive a contribution from the unit operator. In the case of SCFTs, the fact that supersymmetry is unbroken on $S^3\times \R$ is a simple consequence of radial quantization. Indeed, in this case, the spectrum of $H_{\rm susy}$ is identical to the spectrum of $\Delta-\tfrac{1}{2r_3} R$, which has a gap above the unit operator.  
 
Another simple observation is that $J^3_l$ annihilates the vacuum, $J^3_l|{\rm VAC}\rangle=0$. Indeed, $J_l^3$ appears with different signs on the right hand side of $\{Q_1,Q_1^\dagger\}$ and $\{Q_2,Q_2^\dagger\}$. Hence, it must vanish or one of the $Q$'s does not annihilate the vacuum. (Also, if $J^3_l$ were nonzero on the vacuum, the vacuum would not be unique.)

It is useful to focus on the algebra of one specific supercharge, say $Q_1$. We will also rescale $R$ and $J^3_l$ to reabsorb the radius $r_3$ of $S^3$. Then the $Q_1$-algebra reads 
\bea\label{Q1algebra}
\label{algebra} &&\{Q_1,Q_1^\dagger\} \ =\  2\left(H_{\rm susy}- R - 2 J^3_l \right)~,\qquad Q_1^2\ =\ 0~,\nn \\ [2mm]
&& [H_{\rm susy},Q_1]\ =\ [R + 2J^3_l,Q_1] \ = \ 0\ .
\eea
From this we conclude that 
\bea
\langle H_{\rm susy} \rangle & = & \langle R \rangle ~.
\label{HRvac}
\eea
However, the algebra~\eqref{Q1algebra} is invariant under shifting $H_{\rm susy}$ and the $R$-charge by some c-number $\epsilon$ so we cannot yet determine the actual expectation value of $H_{\rm susy}$ in the vacuum. Notice that an equivalent way to express the Ward identity (\ref{HRvac}) is in terms of the vacuum expectation value of $\Delta$, defined in~\eqref{SUSYH}:
\bea
\langle \Delta \rangle & = & \frac{3}{2} \, \langle R \rangle~.
\label{DRvac}
\eea
Equation (\ref{DRvac}) takes the form of the familiar BPS relation.  
Here we interpret it on the cylinder, where both sides have nonzero vacuum expectation values.

We will approach the problem of determining $\langle H_{\rm susy}\rangle$ by reducing the theory on the three-sphere. In this way we get a quantum mechanics (QM) theory with infinitely many degrees of freedom. The theory has four supercharges, $Q_1,Q_2$ and their Hermitian conjugates. The $R$-symmetry group is $SU(2)_l\times U(1)$ and the supercharges furnish the $(2,1)$ representation. $SU(2)_r$ is a global symmetry in the quantum mechanics theory.

The generating functional of $R$-current connected correlators  discussed in (\ref{QMcounterintro}) is specifically 
 given by
\begin{align}\label{QMcounter}
& \mathcal W[A_t] \ =\  \langle R\rangle  \int \diff t  \, A_t \ , 
\end{align}
where $A_t$ is the component along $S^1$ of the four-dimensional $R$-symmetry gauge field. 
 This term is a $1d$ Chern-Simons term fixing the $R$-charge of the vacuum. Given that $\langle H_{\rm susy}\rangle = \langle R\rangle $, this also fixes the Casimir energy.
Since the relation $\langle H_{\rm susy}\rangle = \langle R\rangle $ is a consequence of supersymmetry, the one-dimensional local term \eqref{QMcounter} must be part of a supersymmetric term in a one-dimensional supergravity, obtained by dimensional reduction of $4d$ new minimal supergravity on $S^3$.
In fact, in the specific case of round $S^1\times S^3$ that we are discussing here, it is easy to see that 
a plausible candidate for the completion of the CS term (\ref{QMcounter}) to a 
$1d$ supergravity invariant  is given by 
\be
\mathcal W  \ = \   \langle R\rangle   \int \diff t \left(  A_t  +  \frac{3}{2r_3}\sqrt{|g_{tt}|}   - \frac{3}{2}V_t \right)~,
\ee
where $g_{tt}$ and $V_t$ are simply the components of the respective background fields in four dimensions.  
This expression does not depend on the choice of the parameter $\kappa$ and  correctly reproduces~\eqref{DRvac} in a SCFT. However, 
we leave the systematic study of such terms for the future.

Equation \eqref{QMcounter}  implies that {\it $\langle H_{\rm susy} \rangle$ is physical} ({\it i.e.}\ scheme independent) without relying on the classification of four-derivatives counterterms of~\cite{Assel:2014tba}.  The idea is that if we just studied a quantum mechanics theory of finitely many degrees of freedom, then there would be no meaningful way to compute $\langle H_{\rm susy}\rangle$ itself. 
Indeed, we could always add a counterterm of the form \eqref{QMcounter} with an arbitrary coefficient, shifting the values of $\langle H_{\rm susy}\rangle$ and $\langle R\rangle$ at will, while preserving $\langle H_{\rm susy} \rangle=\langle R\rangle$.\footnote{This freedom is constrained when the $R$-symmetry is compact. In this case, we can only add~\eqref{QMcounter}\ with an integer coefficient. Thus, we can only change the $R$ charge by an integer amount, such that $e^{2\pi i R}=1$ on all the states is retained. So if the $R$-symmetry were compact, then the ambiguity in the vacuum energy would be only by an integer, even in quantum mechanics.} 
So without additional assumptions, there is no possible way to fix $\langle H_{\rm susy}\rangle$ itself at the level of quantum mechanics.
However, it is crucial that in the present study the quantum mechanics theory derives from a four-dimensional local quantum field theory. 
Then, the Chern-Simons counterterm~\eqref{QMcounter}\ would be admissible only if it came from a local term in four dimensions. It is easy to convince oneself that there is no way to derive~\eqref{QMcounter} --- with its specific normalization --- by dimensional reduction of a four-dimensional counterterm on $S^3 \times \mathbb{R}$. Indeed to get the normalization right for the Chern-Simons term, it needs to come from a term in four dimensions of mass dimension four. But then it is straightforward to see that one cannot write anything, {\it regardless of supersymmetry}, that would look like $\int \diff t  A_t$ after integrating over the sphere with radius $r_3$. Therefore, the charge of the vacuum becomes physical and so does the ground state energy by the relation $\langle H_{\rm susy}\rangle=\langle R\rangle$.

Another observation that follows directly from~\eqref{QMcounter} is that {\it the Casimir energy cannot depend on continuous coupling constants (and hence on the RG scale).} This follows from the fact that $\langle H_{\rm susy}\rangle=\langle R\rangle$ and that to compute $\langle R\rangle$ we need to evaluate the coefficient of $\int \diff t A_t$.  If the coefficient of $A_t$ had depended on continuous coupling constants, we could have promoted them to time-dependent fields and lose the gauge invariance under small $R$-symmetry gauge transformations (and this cannot be accounted for by any anomaly). This is similar to the arguments in~\cite{DiPietro:2014bca}\ (and references therein). This conclusion is consistent with the arguments of~\cite{Closset:2013vra,Closset:2014uda}. 

An important consequence of this observation is that it is sufficient to calculate the Casimir energy starting from a free field theory in $4d$. We henceforth assume that such a free point exists in the space of continuous coupling constants. It would be very interesting to generalize our considerations to non-Lagrangian theories. (It is not currently clear to us how to do so.)

To summarize, by considering the supersymmetry algebra and reducing to a quantum mechanical problem, we established that what we need to compute is the coefficient of the generating functional~\eqref{QMcounter} in quantum mechanics. We perform this computation below, after having introduced some notions of supersymmetric quantum mechanics. 

Before coming to these issues, let us comment about the case in which two supercharges are preserved instead of four. This is in fact the generic supersymmetric case, pertinent to various deformations of $S^3$ (these will be discussed in section 3) as well as to other topologies. 
In this case, some of the claims above are valid while some others are not necessarily true. An important difference is that there is no $SU(2)_l$ algebra in general, rather, there is just some $U(1)$ isometry of $M_3$, generated by $J_3$. Thus, a simple argument that $J_3$ vanishes in the vacuum does not exist. The QM algebra, inherited from the $4d$ deformed algebra \cite{Dumitrescu:2012ha,Assel:2014paa}, now takes the form
\eqref{SUSYQMintro}, so that in the vacuum we have $\langle H_{\rm susy} \rangle = \langle \Sigma\rangle$, with $\Sigma$ a QM flavor symmetry. A quantum mechanical term like \eqref{QMcounter} would still exist, with $A_t$ replaced by the gauge field $A^\Sigma_t$ for the flavor symmetry $\Sigma$, and the claims made below eq.~\eqref{QMcounter} still apply. In particular, the vacuum energy is still independent of the renormalization scheme and of the coupling constants. We will use this to compute $E_\mathrm{susy}$ for $\mathcal N=1$ theories with an $R$-symmetry on $M_3\times S^1$, where $M_3$ has $S^3$ topology.


\subsection{Supersymmetric quantum mechanics}

Let us model the situation above, governed by the two supercharges algebra \eqref{Q1algebra}, with  
\bea\label{model}
&&\{Q,Q^\dagger \} \ = \ 2(H-\Sigma)~,\qquad Q^2 \ =\ 0\ , \nn \\ [2mm]
&&[H, Q]\ =\ [\Sigma, Q]\ =\ 0~,
\eea
where $H$ generates time translations, while $\Sigma$ is some Hermitian conserved charge. At the formal level, we can just redefine $H$ by $\Sigma$. However, in order to be able to connect more easily to the reduction over $S^3$, we will keep the algebra in the form~\eqref{Q1algebra}. Similar supersymmetric systems were studied in~\cite{Smilga:2003gu,Ivanov:2013ova}.

We can define two types of multiplets: a {\it chiral multiplet} $(\phi,\psi)$, and a {\it Fermi multiplet} $(\lambda,f)$, where $\phi,f$ are complex and commuting while $\psi,\lambda$ are complex and anti-commuting. These two multiplets have the following  supersymmetry transformations
\bea\label{1dsusyTrans}
\mathrm{chiral}\ : \;\; &&\delta \phi = \sqrt{2} \zeta \psi ~,  \qquad\qquad\qquad\;\, \delta    \psi = - \sqrt{2} i \zeta^\dagger D_t \phi~, \nonumber\\ [2mm]
\mathrm{Fermi}\ :  \;\; &&\delta \lambda = \sqrt{2} \zeta f  + p \sqrt{2} \zeta^\dagger \phi ~,  \qquad \delta   f  = - \sqrt{2} i \zeta^\dagger D_t \lambda - p \sqrt{2} \zeta^\dagger \psi ~,\qquad 
\eea
where on all the fields we define $D_t =\partial_t -i \sigma$, with $\sigma$ the charge of the field under $\Sigma$. The complex parameter $\zeta$ is independent of time and uncharged under $\Sigma$. In the variations of the Fermi multiplet there  appears a parameter $p$. When $p=0$, the chiral and Fermi multiplets are independent of each other. We will refer to each of the decoupled multiplets as ``short.'' When instead $p \neq 0$ the two multiplets form one reducible but indecomposable representation of supersymmetry. Thus, for $p\neq 0$ we call the combined chiral and Fermi multiplets a ``long'' multiplet.

On each component of a multiplet with charge $\sigma$, the transformations \eqref{1dsusyTrans} give
\be
\{\delta_{1},\delta_{2}\} \ = \ -2i\big (\zeta_1^\dagger \zeta_2 + \zeta_2^\dagger \zeta_1 \big) D_t \ ,
\ee
which is consistent with the algebra \eqref{model} when $H$ is represented as $-i\partial_t$.
 
The supersymmetric Lagrangian of a long multiplet takes the form
\bea\label{1dLagrangian}
L & =& |D_t \phi |^2  - i \mu (\phi D_t\phi^\dagger - \phi^\dagger D_t \phi) + i \psi^\dagger D_t \psi  - 2 \mu \psi \psi^\dagger \nonumber\\ [2mm]
   & &+\,\,  i \lambda^\dagger D_t \lambda  + |f|^2 \nonumber\\ [2mm]
    & &-\,\, p^2 |\phi|^2 - p (\lambda \psi^\dagger + \psi \lambda^\dagger)  ~,
\eea
where $\mu$ is an additional free parameter, giving a mass to $\psi$.
For $p=0$, the first and the second lines are the Lagrangians of a free chiral and free Fermi multiplet, respectively, and are separately supersymmetric.\footnote{When $p=0$ an additional term like $\delta(\lambda W(\phi))$ can be introduced, in case the total charge under $\Sigma$ vanishes. We will not need to consider this term.}

We now pass to Hamiltonian formalism and quantize the theory. The canonical momenta are
\be
\Pi_\phi = (D_t +i\mu)\phi^\dagger\ , \qquad \Pi_\psi = -i\psi^\dagger\ , \qquad\Pi_\lambda = -i\lambda^\dagger\ , \qquad \Pi_f = 0\ .
\ee
The canonical (anti-)commutation relations are
\be
[\phi,\Pi_\phi] = i\ , \qquad \{\psi ,\Pi_\psi \} \equiv -i\{\psi ,\psi^\dagger \} = -i  \ , \qquad \{\lambda ,\Pi_\lambda \} \equiv -i\{\lambda ,\lambda^\dagger \} = -i \ , 
\ee
together with their Hermitian conjugates. 

The Hamiltonian reads
\bea\label{Hamiltonian}
H &=& |\Pi_\phi|^2 + i(\mu+\sigma)(\Pi_\phi \phi - \phi^\dagger \Pi_{\phi^\dagger}) + \mu^2 |\phi|^2 + (\sigma+2\mu)\psi\psi^\dagger \nn \\ [2mm]
&& +\, \sigma \lambda\lambda^\dagger \nn \\ [2mm]
&& +\, p^2|\phi|^2 + p(\lambda\psi^\dagger + \psi\lambda^\dagger) + \widetilde\alpha\ ,
\eea
where again when $p=0$ the first line gives the Hamiltonian of a chiral multiplet, while the second line is the Hamiltonian of a Fermi multiplet. The field $f$ has been set to zero by its equation of motion.
 Note that we have introduced a constant $\widetilde\alpha$, parameterizing the usual ordering ambiguity.

In terms of canonical variables, the charge $\Sigma$ reads
\be\label{Sigcan}
\Sigma \ = \ i\sigma \big( \Pi_\phi \phi - \phi^\dagger \Pi_{\phi^\dagger} \big) + \sigma\big(\psi\psi^\dagger + \lambda\lambda^\dagger \big) + \alpha\ ,
\ee
where $\alpha$ parameterizes the ordering ambiguity in this operator. The supercharge is
\be
Q \ = \ \sqrt{2}i\, \psi\big( \Pi_\phi- i \mu \phi^\dagger \big) + \sqrt{2}\,p\,\phi^\dagger \lambda \ ,
\ee
and is free of ordering ambiguities. Evaluating $\{Q,Q^\dagger\}$ we find that~\eqref{model} is upheld  provided we take 
\be\label{solTiAlpha}
\widetilde\alpha \ =\ \alpha -2\mu\ .
\ee Hence supersymmetry fixes the ordering ambiguity in $H-\Sigma$. Of course, after having solved for $\widetilde\alpha$ we still have the freedom to shift $H$ and $\Sigma$ by an equal amount, corresponding to the remaining parameter $\alpha$. Without additional assumptions, this freedom would have remained in the framework of ordinary quantum mechanics in one dimension.

In order to explain how to fix the ordering ambiguity that is left, it is useful to recall that we are computing the coefficient of a CS term in the low-energy $1d$ effective action. This term takes the form
\be\label{1dCSterm}
k \int \diff t\,  A^\Sigma_t \ ,
\ee
where $A^\Sigma_t$ is the background gauge field associated to the charge $\Sigma$. A single fermion of mass $M$ and charge $q$ shifts the coefficient of the Chern-Simons term by $\frac{q}{2} {\rm sgn}(M)$~\cite{Elitzur:1985xj}. We can think about that as if we are starting from some theory in the UV with Chern-Simons coefficient $k_{\rm uv}$ and  then we integrate out the massive fermion leading to a Chern-Simons coefficient in the infrared $k_{\rm ir}$ (this interpretation was elaborated upon in~\cite{Closset:2012vp})\footnote{A simple way to derive (\ref{differenceCS}) is as follows. First, from dimensional analysis and the fact that $M$ and $k$ are odd under charge conjugation we infer  
\be\label{difgeneral}
k_{\rm ir} - k_{\rm uv} \ = \  x \, {\rm sgn}(M)  \ ,
\ee
where $x$ is a coefficient, independent of $M$.
To fix $x$ we can consider a free fermion with mass $M$ and charge $q$ with a 
constant background gauge field  $A^\Sigma_t$. This has Hamiltonian $H = (M  + q A^\Sigma_t) (\psi \psi^\dagger +\hat \alpha)$, where $\hat \alpha$ is an arbitrary  ordering constant.
The partition function is given by 
\bea
Z & = & e^{-\beta (M + q A^\Sigma_t )\hat \alpha}\left( 1 + e^{-\beta (M+ q A^\Sigma_t )} \right) ~.
\label{Zfree}
\eea
The idea now is that we can keep the ultraviolet fixed and consider two different RG flows, one with positive $M$ and one with negative $M$. By subtracting the resulting Chern-Simons terms in the infrared (which we will read out from the charge of the vacuum), we will find $2x$.
If $M>0$ then taking $M\to \infty$ we can read off the CS term ({\it i.e.}\ charge) in the IR to be $q \hat \alpha \int \diff t A^\Sigma_t $. On the other hand, if $M<0$ we read out the CS term in the IR by taking the limit $M\rightarrow -\infty$ and we find 
$q (\hat \alpha +1) \int \diff t A^\Sigma_t $. Subtracting these yields $2x=- q$.}
\be\label{differenceCS}
k_{\rm ir} - k_{\rm uv} \ = \ -\frac{q}{2}\, {\rm sgn}(M)  \ .
\ee

From the point of view of the quantum mechanics, the arbitrariness in the charge of the vacuum corresponds to the arbitrariness in the UV coefficient $k_{\rm uv}$. However, our theory arises from a higher-dimensional model. As already observed, it is easy to convince oneself that a term like \eqref{1dCSterm} cannot be generated by dimensional reduction of a four-dimensional local term. So we must take 
\be\label{crit}
k_{\rm uv} = 0~,
\ee 
\emph{i.e.}\ no Chern-Simons contact term in the UV generating functional. This key requirement fixes the ordering ambiguity in $H$. Together with \eqref{differenceCS}, this implies that multiplets containing pairs of fermions with masses of opposite sign do not contribute to the Casimir energy. We will see below that as long as the Hamiltonian is bounded from below, a long multiplet necessarily contains fermions with masses of opposite sign. As a result, the choice of the ordering coefficient must be such that $H$ and $\Sigma$ vanish in the ground state of a long multiplet. This leads to the conclusion that the correct choice of the ordering constant is \be\label{ordering} \alpha = - 2 \sigma~.\ee We will use this choice in the following and one can verify that in all cases the results are consistent with~\eqref{crit}.
Incidentally, it turns out that~\eqref{ordering} also corresponds to Weyl ordering for the Hamiltonian.\footnote{This explains why our final result is identical to that of~\cite{Lorenzen:2014pna} for the VEV of $H$. But, unlike~\cite{Lorenzen:2014pna}, our result for the VEV of $\Sigma$ in the vacuum manifestly 
respects the BPS condition $H=\Sigma$.\label{foot:compareDJ}}

\subsection{Spectrum of the Hamiltonian}
\label{SpectrumH}

We now study the spectrum of the Hamiltonian and determine the vacuum state. 

\subsubsection*{Long multiplet}

Let us start from the bosonic sector of \eqref{Hamiltonian}:
\be
H_{\rm bosonic} \ = \ |\Pi_\phi|^2+i(\mu+\sigma) (\Pi_\phi\phi-\phi^\dagger\Pi_{\phi^\dagger})+(\mu^2+p^2)|\phi^2|-\mu-\sigma\ ,
\ee
where we have included half of the ordering constant appearing there (the other half will enter in the fermionic sector). This ensures Weyl ordering.
We can introduce creation operators $a^\dagger$, $b^\dagger$ and annihilation operators $a$, $b$ via
\be
\phi \ =\ \frac{(\mu^2 + p^2)^{-1/4}}{\sqrt 2}\big(a+ b^\dagger\big)\ , \qquad \Pi_\phi \ =\ \frac{i(\mu^2 + p^2)^{1/4}}{\sqrt 2}\big(a^\dagger - b\big)\ .
\ee
The canonical commutation relations between $\phi$ and $\Pi_\phi$ (and their Hermitian conjugates) imply that these satisfy $[a,a^\dagger]=[b,b^\dagger]=1$, $[a,b]=[a^\dagger,b]=[a,b^\dagger]=[a^\dagger,b^\dagger]=0$. Then the bosonic Hamiltonian
can be written as
\bea
H_{\rm bosonic} &=&  \sqrt{\mu^2+p^2}\big(a^\dagger a + b^\dagger b + 1\big) + (\sigma+ \mu)\big( b^\dagger b - a^\dagger a \big)\nn \\ [2mm]
&=& \frac{1}{2}\sqrt{\mu^2+p^2}\big(\{a, a^\dagger\} + \{b, b^\dagger\}\big)  + \frac{1}{2}(\sigma+ \mu)\big(\{b, b^\dagger\} - \{a, a^\dagger\} \big)\ ,
\eea
where in the second line we have emphasized that $H_{\rm bosonic}$ is Weyl ordered.
The state annihilated by $a$ and $b$ has energy $\sqrt{\mu^2+ p^2}\,$. Acting on this with $(a^\dagger)^m(b^\dagger)^n$ (with $m,n$ positive integers) we obtain a state with energy
\be
H_{{\rm bosonic}}(m,n) \ =\  \sqrt{\mu^2 + p^2} + m\big( \sqrt{\mu^2 + p^2} -\mu - \sigma\big) + n \big(\sqrt{\mu^2 + p^2} + \mu + \sigma\big)\ .
\ee
We see that in order for the Hamiltonian to have a spectrum that is bounded from below we need to assume $\sqrt{\mu^2+p^2}> |\mu+\sigma|$.\footnote{Allowing for $\sqrt{\mu^2+p^2} = |\mu+\sigma|$ yields a Hamiltonian bounded from below but introduces a degenerate vacuum. Let us discard this case.} Hence the state of minimum energy in the bosonic sector is the one with $m=n=0$.

Next we address the fermionic sector. The Hamiltonian reads
\bea
H_{\rm fermionic} &=& p(\lambda\psi^\dagger+\psi\lambda^\dagger)+(2\mu+\sigma)\psi\psi^\dagger+\sigma\lambda\lambda^\dagger -\mu - \sigma \nn \\ [2mm]   
&=& \big(  \psi \quad \lambda\big)\left(\begin{array}{cc} 2\mu+\sigma &p \\ p &\sigma \end{array}\right)\left(\begin{array}{c} \!\psi^\dagger\! \\ \!\lambda^\dagger\! \end{array}\right) -\mu - \sigma\ ,
\eea
where we have kept the ordering constant that ensures Weyl ordering.
We can make a unitary $U(2)$ rotation to diagonalize the above matrix. 
This preserves the anti-commutation relations. 
The eigenvalues are
\be
\label{eigen} 
x_{\pm}\ =\ \mu+\sigma \pm\sqrt{\mu^2 +p^2}\ .
\ee
Denoting the eigenvectors $u_+,u_-, u_+^\dagger, u_-^\dagger$, the Hamiltonian is thus 
\bea
H_{\rm fermionic} &=& x_+ u_+u_+^\dagger+x_- u_-u_-^\dagger -\mu - \sigma \nn \\ [2mm]
&=& \frac{x_+}{2}\, [u_+,u_+^\dagger] + \frac{x_-}{2}\, [u_-,u_-^\dagger]\ ,
\eea
with $\{u_{\pm},u_{\pm}^\dagger\}=1$. 
The charge operator $\Sigma$ takes the form
\bea
\Sigma_{\rm fermionic}  &=& \sigma\big(u_+u_+^\dagger+u_-u_-^\dagger - 1\big)  \nn \\ [2mm]
&=& \sigma\, [u_+,u_+^\dagger] +\sigma\, [u_-,u_-^\dagger]\ .
\eea
Starting with the state $|0\rangle$ which is annihilated by both $u_{\pm}^\dagger$, we can act with $u_-$, $u_+$ or $u_-u_+$.
The spectrum therefore consists of four states with the following energy and charge:
\be
\begin{array}{c|cccc}
{\rm state} & \ |0\rangle      & 	u_-|0\rangle    & u_+|0\rangle   &   u_+u_-|0\rangle   \\ [2mm]
{\rm energy} &\ -\mu - \sigma\  & \ -\sqrt{\mu^2 +p^2}\;\  & \sqrt{\mu^2 +p^2}\;\    & \ \mu + \sigma \\ [2mm]
{\rm charge}  &\  -\sigma  & 0 & 0  &\sigma
\end{array}
\ee
Since we assumed $\sqrt{\mu^2+p^2}> |\mu+\sigma|$, the state of lowest energy is $u_-|0\rangle$.

We now combine the information obtained studying the bosonic and fermionic sectors of the Hamiltonian and identify a state with minimum energy that respects supersymmetry.
Adding $H_{\rm bosonic}$ and $H_{\rm fermionic}$, the complete Hamiltonian is
\bea
H &=& \sqrt{\mu^2+p^2}\big(a^\dagger a + b^\dagger b + 1\big) + (\sigma+ \mu)\big(b^\dagger b - a^\dagger a\big) \nn \\ [2mm]
&& +\, x_+ u_+u_+^\dagger+x_- u_-u_-^\dagger - \mu - \sigma \ .
\eea
One can also check that the full charge operator reads
\be
\Sigma \ = \ \sigma\, \big( b^\dagger b - a^\dagger a + u_+u_+^\dagger+u_-u_-^\dagger  -1 \big)\ .
\ee
From the discussion above, the state with minimum energy is clearly
\be
|{\rm VAC}\rangle\ \equiv\ |m=0,n=0,x_-\rangle~,
\ee
where $m=0$, $n=0$ indicates that no bosonic oscillators are excited, and by $x_-$ we mean that we excite one fermionic oscillator with eigenvalue $x_-$. Its total energy is
\be
H\ = \ \sqrt{\mu^2+ p^2} -\sqrt{\mu^2 +p^2} \ =\ 0\ , 
\ee
and thus vanishes due to an exact cancellation between the bosonic and the fermionic contributions.
Since we have just one fermionic oscillator the charge is $\Sigma = 0\,,$
hence the relation $(H-\Sigma)|{\rm VAC}\rangle = 0$ is satisfied and supersymmetry is unbroken in the vacuum, as expected.

We conclude that the long multiplets yield a vanishing contribution to the vacuum energy and charge:
\be
\langle H_{\rm long} \rangle \ = \ \langle \Sigma_{\rm long} \rangle \ = \ 0\ .
\ee
Note that this is a consequence of our choice of ordering constant, and as argued at the end of the previous subsection this is the correct choice for a quantum mechanics arising from a higher-dimensional theory.

If we had a theory of long multiplets only, the vacuum energy would just be zero. However, if short multiplets are also present, this is not the case, as we now show.

\subsubsection*{Fermi multiplet}

Consider the Fermi multiplet. Then the supercharge identically vanishes. The Hamiltonian and the charge generator take the same form,
\be
H_{\rm Fermi} \ = \ \Sigma_{\rm Fermi} \ = \ \sigma \Big(\lambda \lambda^\dagger - \frac{1}{2}\Big) \ .
\ee
The only two states have energy $-\frac{1}{2}\sigma$ and $+\frac{1}{2}\sigma$. The contribution of a Fermi multiplet to the vacuum energy and charge is thus
\be\label{contribHfermi}
\langle H_{\rm Fermi} \rangle \ = \ \langle \Sigma_{\rm Fermi} \rangle \ = \ - \frac{|\sigma|}{2}\ .
\ee

\subsubsection*{Chiral multiplet}

The bosonic sector of the chiral multiplet can be treated as we did for the long multiplet, setting $p=0$. The full Hamiltonian and charge operator can thus be written as
\be
H_{\rm chiral} \ =\ |\mu|\big(a^\dagger a + b^\dagger b + 1\big) + (\sigma+ \mu)\big(b^\dagger b - a^\dagger a\big) + (2\mu + \sigma)\psi\psi^\dagger -\mu - \frac{\sigma}{2}\ , 
\ee
\be
\Sigma_{\rm chiral} \ = \ \sigma\left( b^\dagger b - a^\dagger a \right) + \sigma\psi\psi^\dagger  -\frac{1}{2}\sigma\ .
\ee
Since $p =0$, the condition for the Hamiltonian to be bound from below becomes
\be\label{inequality}
|\mu|> |\mu+\sigma|\ .
\ee 
In the vacuum all bosonic oscillators are zero. Then we have two possible states:
\begin{enumerate}
\item the state annihilated by $\psi^\dagger$, with 
$H = |\mu| - \mu - \frac{1}{2}\sigma$ and $\Sigma = -\frac{1}{2}\sigma\,;$
\item the state with an oscillator $\psi$ excited, with
$H= |\mu| + \mu + \frac{1}{2}\sigma$ and $\Sigma = +\frac{1}{2}\sigma$\ .
\end{enumerate}
Which state has minimum energy depends on the values of $\mu$ and $\sigma$. Note that \eqref{inequality} requires $\mu$ and $\sigma$ to have opposite signs. If $\mu>0$, $\sigma<0$, then \eqref{inequality} implies $-2\mu<\sigma<0$, and the state number 1 has minimum energy $H = -\frac 12 \sigma$; since $H=\Sigma$, this state is supersymmetric, while the state 2 is non-supersymmetric. Conversely, if $\mu <0$ and $\sigma>0$, then from \eqref{inequality} we deduce $0<\sigma<-2\mu$, hence the state number 1 now has higher energy and the state 2 is the supersymmetric vacuum, with $H = \frac 12 \sigma$. 

Thus, a chiral multiplet contributes to the vacuum energy and charge as
\be\label{contribHchiral}
\langle H_{\rm chiral} \rangle \ = \ \langle \Sigma_{\rm chiral} \rangle \ = \  \frac{|\sigma|}{2}\ .
\ee

\medskip

In conclusion, the analysis in supersymmetric quantum mechanics establishes that a long multiplet yields a vanishing contribution to the vacuum energy and charge, that a Fermi multiplet contributes as in \eqref{contribHfermi} while a chiral multiplet contributes as in~\eqref{contribHchiral}.


\subsection{Dimensional reduction of a 4d chiral multiplet}\label{ReductionRoundS3}

Consider a free four-dimensional chiral multiplet $(\phi,\psi_\alpha,F)$ on $S^3 \times\mathbb R $. The Lagrangian and supersymmetry transformations can be found in \cite{Festuccia:2011ws}. The only parameter appearing in the Lagrangian is the charge $r$ under the background $R$-symmetry gauge field. Here we will restrict to $0< r \leq 2$.\footnote{Outside this range there are complications, see \cite{Gerchkovitz:2013zra}. Here we see additional complications, for example, the cancelation previously discussed for long multiplets would fail.} This range is compatible with the inequalities mentioned in the previous subsection, ensuring that the spectrum of the Hamiltonian is bounded from below. Expanding in appropriate spherical harmonics, the chiral multiplet reduces to a one-dimensional theory with infinitely many fields. These organize in one-dimensional multiplets with different values of the parameters $\mu,p,\sigma$ introduced above. Some have $p\neq 0$ and are thus long multiplets, while some others have $p=0$ and are thus short multiplets, either chiral or Fermi. 

More explicitly, we can expand the scalars in spherical harmonics $Y_{\ell,m,n}$ transforming in representations $(\frac{\ell}{2},\frac{\ell}{2})$ of $SU(2)_l\times SU(2)_r$. The quantum number $\ell$ is a non-negative integer. For a fixed $\ell$, the quantum numbers $m,n$ of the scalar harmonic $Y_{\ell,m,n}$ range in $-\frac{\ell}{2} \leq m,n \leq  \frac{\ell}{2}\,$. 
So we can write
\bea
\phi & = & \sum_{\ell,m,n} \phi_{\ell,m,n} Y_{\ell,m,n} ~,
\eea
and similarly for the auxiliary field $F$. The fermionic field $\psi_\alpha$ can be expanded in spinorial harmonics. A review of spinor spherical harmonics on $S^3$ can be found in~\cite{Gerchkovitz:2013zra,Lorenzen:2014pna}. A single $4d$ fermion reduces to two infinite series of $1d$ fermions furnishing the representation $\sum_l (\frac{\ell-1}{2},\frac{\ell}{2})\oplus(\frac{\ell+1}{2},\frac{\ell}{2})$ of $SU(2)_l\times SU(2)_r\,$.\footnote{The symmetry between left and right is broken by the choice of spin bundle.}

Integrating over $S^3$ and using the orthonormality of the spherical harmonics, the action of a four-dimensional chiral multiplet gives rise to a one-dimensional action for an infinite number of fields. These arrange in multiplets of supersymmetric quantum mechanics labeled by $\ell,m,n$, and one can check that the Lagrangian of each of these multiplets takes the form~\eqref{1dLagrangian}. 
Here we do not need to present all details of the reduction. All we need to know is how the $R$-charge $r$ and the quantum numbers $\ell,m,n$ map into the parameters $\sigma,p,\mu$ entering in~\eqref{1dLagrangian} and characterizing each multiplet in supersymmetric quantum mechanics. 
Actually, the discussion in subsection~\ref{SpectrumH} shows that for the purpose of determining the vacuum energy we just need to know when a multiplet is shortened (namely when $p=0$), if it is a chiral or a Fermi multiplet, and what is the value of its charge $\sigma$. 

By comparing the four-dimensional algebra \eqref{algebra} with \eqref{model}, 
we deduce that we must identify (restoring the $S^3$ radius $r_3$) $\Sigma \ = \ \frac{1}{r_3}(R + 2 J^3_l)$, 
and therefore
\bea
\sigma & = &  \frac{1}{r_3}( r + 2m ) \ .
\eea
Moreover,  reducing the four-dimensional Lagrangian to one dimension, 
one finds\footnote{More generally, one could easily restore the dependence on the parameter $\kappa$.  This affects only $\mu$ but not $p^2$ and $\sigma$.
In the notation of \cite{Lorenzen:2014pna} one finds that the parameter $\mu$ is related with the parameters in the four-dimensional 
Lagrangian as $r_3 \mu  = - 2m -\frac{3}{2}r -\kappa (\frac{3}{2}r-\epsilon ) $.}
\bea
p^2 & = & \frac{1}{r_3^2}(\ell - 2m )(\ell + 2 + 2m)~,\nonumber\\
\mu & = & -\frac{1}{r_3}(2m +1) ~,
\eea
hence the shortening condition $p=0$ is satisfied if and only if $m = \ell/2$ or $m = -\ell/2 -1$. In the former case a chiral multiplet is obtained, with charge $\sigma = \frac{1}{r_3}(\ell+r)$. In the latter case a Fermi multiplet is obtained, with charge $\sigma = -\frac{1}{r_3}(\ell +2-r)$.
Recalling~\eqref{contribHfermi}, \eqref{contribHchiral} we conclude that the respective contribution to the vacuum energy is: 
\be
\begin{array}{lll}\mathrm{chiral} &\big(m=\frac{\ell}{2}\big): & \quad\langle H_\mathrm{chiral}\rangle   \  = \  \frac{1}{2r_3} (\ell +r) \ ,  \\ [3mm]
\mathrm{Fermi} & (m=-\frac{\ell}{2}-1): & \quad\langle H_\mathrm{Fermi} \rangle  \ = \ -  \frac{1}{2r_3} (\ell +2- r) \ . \end{array}
\ee
The expectation value of the Hamiltonian is obtained by adding up the contributions of all chiral and Fermi multiplets:
\bea\label{CasimirRound}
\langle  H_{\rm susy} \rangle  & = &  \sum_\mathrm{chiral} \langle H_{\rm chiral}\rangle + \sum_\mathrm{Fermi} \langle H_{\rm Fermi}\rangle   \nn \\ [2mm]
& = &  \sum_{\ell \geq 0} \frac{1}{2r_3}  (\ell+1) (\ell +r)  -  \sum_{\ell \geq0} \frac{1}{2r_3}(\ell+1) (\ell +2- r) \ ,
\eea
where the $(\ell+1)$ factor comes from the degeneracy associated with $SU(2)_r$.

To regularize the sum, we dress the terms in the sum with some decreasing weights. To do this in a supersymmetric fashion, we can decompose $ H $ as a sum of Hamiltonians acting on the Hilbert space of a single free $1d$ multiplet
\be
H_{\rm susy} \ = \ \sum_{\ell, m, n} H_{\ell, m, n} \, ,
\ee
and regularize the sum with a function of the $H_{\ell, m, n}$ operators, for instance
\be
H_{\rm susy} \ = \ \sum_{\ell, m, n} H_{\ell, m, n} e^{ -  2 \,  t \, r_3  | H_{\ell, m, n}| } \ ,
\ee
with $t$ a positive number. This yields
\be
\langle  H_{\rm susy} \rangle \ = \  \sum_{\ell \geq 0}   \frac{1}{2r_3}(\ell+1) (\ell +r) e^{- t (\ell +r)} -  \sum_{\ell \geq0} \frac{1}{2r_3} (\ell+1) (\ell +2- r) e^{- t (\ell +2-r)}  \ .
\ee
Taking the small $t$ limit and dropping the diverging term in $t^{-2}$,\footnote{The diverging term can be associated to the four-dimensional Einstein-Hilbert counterterm. We will discuss this more in appendix~C.} we obtain a regularized result for the vacuum energy that, after recalling the trace anomaly coefficients~\cite{Anselmi:1997am}
\be
a = \frac{3}{32} \left[  3(r-1)^3 - (r-1)   \, \right] ~, \qquad\quad
c = \frac{1}{32}  \left[   9(r-1)^3 - 5 (r-1) \right]~,
\label{AnomCoeff}
\ee
reads
\be
E_{\rm susy} \ = \ \langle  H_{\rm susy} \rangle \ = \ \frac{4}{27 r_3}(a+3c)\ .
\ee
This is the result advertised in eq.~\eqref{secondesusy} of the introduction.

One could consider a supersymmetric regularization with a different function $f (t H_{\ell, m, n})$ of the $H_{\ell, m, n}$ operators. It can be shown, using an Euler-MacLaurin expansion (see appendix \ref{app:1LoopDet} for a related application) that for all smooth functions $f$ such that $f(0)=1$ (and such that the series converges), one obtains the same result for the finite piece in the small $t$ expansion. This is in agreement with the fact that the supersymmetric Casimir energy is unambiguous.  The regularization using the Hurwitz zeta function~\cite{Lorenzen:2014pna} also reproduces the same result.

It is possible to contrast our results with several previous works in which localization techniques on $S^3\times S^1$ were utilized. Comparing with~\cite{Assel:2014paa} (see also  \cite{Closset:2013sxa} and \cite{Nishioka:2014zpa} where similar localization techniques are used in other topologies), one finds agreement regarding the vacuum energy. However, as we will briefly discuss in appendix~\ref{app:1LoopDet}, the regularization scheme of~\cite{Assel:2014paa}
in fact does not preserve supersymmetry, as it violates certain SUSY Ward identities in the small circle limit. Our result for the vacuum energy also agrees with that of~\cite{Lorenzen:2014pna} (this method is a Hamiltonian version of~\cite{Kim:2012ava}), but as mentioned in footnote~\ref{foot:compareDJ} the result of~\cite{Lorenzen:2014pna} does not preserve some SUSY Ward identities as well.


\section{Supersymmetric Casimir energy on a deformed three-sphere}\label{DeformedS3}

In this section we study the chiral multiplet on a supersymmetric $S^3 \times S^1$ background with more general metric and complex structure. We will use results known from the computation of the partition function on these spaces, based on localization, to implement the Hamiltonian approach to the evaluation of the Casimir energy. 


\subsection{Shortening conditions on chiral multiplets}\label{ChiralMultRevisited}

We start by reviewing some results of \cite{Closset:2013sxa}, where the fermionic degrees of freedom in the chiral multiplet were conveniently redefined. This made it particularly easy to show that in backgrounds preserving two supercharges of opposite $R$-charge, the modes contributing to the partition function solve first-order differential equations that can be interpreted as shortening conditions. 

We work in Euclidean signature and follow the conventions of~\cite{Assel:2014paa}.\footnote{We could also work in Lorentzian signature, building on the results of~\cite{Cassani:2012ri}. One reason for choosing the Euclidean signature is that it allows to make contact with the computation of the partition function via localization.} We assume the existence of at least one positive-chirality spinor $\zeta_\alpha$ and one negative-chirality spinor $\ti\zeta^{\dot\alpha}$ satisfying the Killing spinor equations
\bea
\left( \nabla_{\mu} - i A_\mu \right) \zeta  + i V_{\mu} \zeta + i V^{\nu} \sigma_{\mu\nu} \zeta \!& = &\! 0  \ , 
\nn \\ [2mm]
\left( \nabla_{\mu} + i A_\mu \right) \ti\zeta  - i V_{\mu} \ti\zeta - i V^{\nu} \ti\sigma_{\mu\nu} \ti\zeta \!& = &\! 0 \ .\label{KeqnZetaTiZeta}
\eea
These independent equations are the Euclidean version of the supersymmetry condition given in~\eqref{ksei}. Note that $\zeta$ has $R$-charge $+1$ while $\ti\zeta$ has $R$-charge $-1$. In Euclidean signature the chiral multiplet is made of $(\phi,\psi_\alpha,F)$ with $R$-charge $(r,r-1,r-2)$ and of the independent fields $(\ti \phi,\ti\psi^{\dot \alpha},\ti F)$ with $R$-charge $(-r,-r+1,-r+2)$.
The supersymmetry transformations are
\bea
 \delta \phi &=&  \sqrt{2}\, \zeta \psi \ ,\qquad\qquad \qquad\qquad\qquad\quad\; \delta \ti\phi  \ =\ \sqrt{2}\, \ti\zeta \ti\psi \;, \nn\\ [2mm]
 \delta \psi  &=&  \sqrt{2}\, F \zeta + \sqrt{2}i \sigma^{\mu}\ti\zeta  D_{\mu} \phi \ , \qquad\qquad\;\;  \delta \ti\psi  \ =\  \sqrt{2}\, \ti F \ti\zeta + \sqrt{2}i \,\ti\sigma^{\mu}\zeta D_{\mu} \ti\phi \;, \qquad\nn \\ [2mm]
 \delta F  &=& \sqrt{2}i\,  D_{\mu}\big(\ti\zeta\, \ti\sigma^{\mu}\psi\big)\ , \qquad\qquad\qquad\quad\,  \delta \ti F  \ =\ \sqrt{2}i\,  D_{\mu}\big(\zeta \sigma^{\mu}\ti\psi\big) \ , \label{ChiTransfo}
\eea
where on a field of $R$-charge $q$ the covariant derivative $D_\mu$ is defined as
\be
D_\mu \ = \ \nabla_\mu - i q A_\mu\ ,
\ee
with $\nabla_\mu$ the Levi-Civita connection.
A supersymmetric Lagrangian is
\bea
\scL  &=& D_{\mu}\ti \phi D^{\mu} \phi   + V^{\mu} \big( i  D_{\mu}\ti \phi\,\phi - i \ti \phi D_{\mu}\phi \big)+ \frac{r}{4} \left( R + 6 V_{\mu}V^{\mu} \right) \ti\phi \phi     - \ti F F
   \nn\\
 &&  +\,      i \ti \psi \,\ti\sigma^{\mu}D_{\mu}\psi  + \frac{1}{2} V^\mu \ti\psi\, \ti \sigma_{\mu} \psi
 \ ,\label{Lagrangians:chi}
\eea
where $R$ is the Ricci scalar on the four-manifold.

Following \cite{Closset:2013sxa} (see section 5 therein), we decompose the fermion fields $\psi$, $\ti \psi$ in anti-commuting scalars as\footnote{Note that $\zeta_\alpha$ and $\varepsilon_{\alpha\beta}(\zeta^\dagger)^{\beta}$ form a basis of chiral spinors.}
\be
B \ = \ \frac{1}{\sqrt 2}\frac{\zeta^\dagger \psi}{|\zeta|^2}\ , \qquad\;  C \ =\ \sqrt 2\,\zeta\psi \qquad\, \Leftrightarrow \qquad \psi_\alpha \ = \ \sqrt 2\,\zeta_\alpha B - \frac{1}{\sqrt 2}\frac{\varepsilon_{\alpha\beta}\zeta^{\dagger\,\beta}}{|\zeta|^2}C \, ,
\ee
\be
\ti B \ = \ \frac{1}{\sqrt 2}\frac{\ti\zeta^\dagger \ti \psi}{|\ti\zeta|^2}\ , \qquad \ti C \ = \ \sqrt 2\,\ti\zeta\ti\psi  \qquad \Leftrightarrow \qquad \ti\psi^{\dot\alpha} \ = \ \sqrt 2\,\ti\zeta^{\dot\alpha} \ti B - \frac{1}{\sqrt 2}\frac{\varepsilon^{\dot\alpha\dot\beta}\,\ti\zeta^{\,\dagger}_{\dot \beta} }{|\ti\zeta|^2}\ti C\ .
\ee
Note that $B$ has $R$-charge $r-2$ while $C$ has $R$-charge $r$. Similarly, $\ti B$ has $R$-charge $-r+2$ while $\ti C$ has $R$-charge $-r$.
We also introduce the complex vectors
\be\label{defvectors}
K^\mu =  \ti\zeta\,\ti\sigma^\mu \zeta\ , \qquad \overline K{}^\mu = \frac{\ti\zeta^\dagger\ti\sigma^\mu \zeta^\dagger}{4|\zeta|^2|\ti\zeta|^2}  \ , \qquad  Y^\mu = \frac{\ti\zeta^\dagger\ti\sigma^{\mu}\zeta}{2|\zeta|^2} \ , \qquad \overline Y{}^\mu  =  -\frac{\ti\zeta\,\ti\sigma^{\mu}\zeta^\dagger}{2|\zeta|^2}\ ,
\ee
which define a complex frame. These satisfy $K_\mu \overline K{}^\mu = Y_\mu \overline Y{}^\mu = \frac 12$, with all other contractions vanishing. The vectors
$K$, $\overline K$ have vanishing $R$-charge, while $Y$ has $R$-charge $+2$ and $\overline Y$ has $R$-charge $-2$. They satisfy 
\be
\nabla_{(\mu} K_{\nu )}=0\ ,\qquad \nabla_\mu \overline K{}^\mu =0\ , \qquad  D_\mu Y^\mu =0\ , \qquad D_\mu \overline Y{}^\mu = 0\ ,
\ee
hence $K$ is a (complex) Killing vector.
Finally, we define the differential operators
\be\label{DiffOp}
\hat{\mathcal L}_K = K^\mu D_\mu\ , \qquad \hat{\mathcal L}_{\overline K} = \overline K{}^\mu D_\mu\ , \qquad \hat{\mathcal L}_Y = Y^\mu D_\mu\ , \qquad \hat{\mathcal L}_{\overline Y} = \overline Y{}^\mu D_\mu\ .
\ee

In the new variables the supersymmetry transformations \eqref{ChiTransfo} take the form (here we distinguish between $\delta\equiv\delta_{\zeta}\,$ and $\ti\delta\equiv\delta_{\ti\zeta}\,$):
\bea
 \delta \phi &=&   C \ , \qquad \qquad\qquad  \ti\delta \phi \ =\  0 \ ,\nn\\ [2mm]
 \delta C  &=&  0 \ , \qquad\qquad\qquad\, \ti\delta C  \ =\  2i\, \hat{\mathcal L}_K \phi\ , \nn\\ [2mm]
 \delta B  &=&   F  \ , \qquad\qquad \quad\;\;\,\, \ti\delta B  \ = \  - 2i\,\hat{\mathcal L}_{\overline Y} \, \phi\ , \nn\\ [2mm]
 \delta F  &=& 0 \ , \qquad \qquad\qquad \,\ti\delta F  \ =\  2i\,  \big( \hat{\mathcal L}_K B + \hat{\mathcal L}_{\overline Y}  C \big)
 \ , \label{ChiTransfoTwisted}
\eea
with similar transformations for the tilded fields. 
The Lagrangian \eqref{Lagrangians:chi} can then be rewritten as
\bea\label{ChiralLtwisted}
\mathcal L &=&  4\, \scL_{\overline K} \ti \phi\, \hat{\scL}_K \phi + 4\, \scL_{Y} \ti \phi \,\hat{\scL}_{\overline Y} \phi  + i \kappa\big( \hat{\scL}_K \ti \phi\, \phi - \ti \phi \hat{\scL}_K \phi \big) - \ti F  F \nn \\ [2mm]
&& + \,2i \ti B \hat{\scL}_K B + 2 i\,\ti C \hat{\scL}_{\overline K}C + 2 i\,\ti B \hat{\scL}_{\overline Y}C -  2 i\,\ti C \hat{\scL}_{Y}B - \kappa \ti C C \ .
\eea
Here, $\kappa$ is a function describing a redundancy in the choice of the background fields~\cite{Dumitrescu:2012ha}; it will play no important role for us as it drops from the final answer.

In~\cite{Closset:2013sxa} it was showed that in a background with two supercharges $\zeta$, $\ti \zeta$, the partition function of a chiral multiplet
reduces to
\be\label{reducedZ}
Z \ = \ \frac{\prod \lambda^B}{\prod \lambda^\phi}\ ,
\ee
where $\lambda^B$, $\lambda^\phi$ are eigenvalues determined by the first-order differential conditions
\be
\hat{\mathcal L}_Y B \ = \ 0 \ , \qquad \qquad i \hat{\mathcal L}_K B \ = \ \lambda^B B \ ,
\label{1stOrderConditionsB}
\ee
and
\be
\hat{\mathcal L}_{\overline Y}\, \phi \ = \ 0 \ , \qquad \qquad i \hat{\mathcal L}_K \phi \ = \ \lambda^\phi \phi\ .\label{1stOrderConditionsphi}
\ee
The equations on the left can be read as shortening conditions for the chiral multiplet. Indeed, $\hat{\mathcal L}_Y B  = 0$ allows to set $C$ (and $\phi$) to zero consistently with its equation of motion. Similarly, $\hat{\mathcal L}_{\overline Y}\, \phi = 0$ permits to set $B$ to zero consistently with its supersymmetry transformation in~\eqref{ChiTransfoTwisted}. In order to set $B=0$ consistently with its equation of motion one also needs $\hat{\mathcal L}_{\overline Y}C = 0$; this condition is also needed to set $F=0$ respecting its supersymmetry variation. Similar considerations apply to the tilded fields. (In~\cite{Closset:2013sxa}, the latter were related to the untilded fields via Hermitian conjugation).

These shortenings are better understood by reducing to lower dimensions, where  $(B, F)$ and $(\phi,C)$ define irreducible representations of supersymmetry. In~\cite{Closset:2013sxa}, this was done in an $S^2 \times \mathbb{T}^2$ background by expanding in monopole harmonics on $S^2$ and noting that the aforementioned  equations correspond to shortenings of the $(0,2)$ supersymmetry multiplets on $\mathbb{T}^2$.
Therefore only modes in shortened multiplets contribute to the one-loop determinant of the chiral multiplet partition function. The long multiplets lead to paired bosonic and fermionic eigenmodes and hence yield a trivial contribution. 

Similarly, in the following we consider $S^3 \times S^1$ with a general metric and reduce on $S^3$ to supersymmetric quantum mechanics. This allows to make contact with the approach to the computation of the Casimir energy developed in section~\ref{sec:SusyVacE}.


\subsection{Reduction on deformed three-sphere}\label{sectionWithb1b2}

We consider the class of supersymmetric backgrounds studied in \cite{Assel:2014paa}. These are complex manifolds with the topology of $S^3 \times S^1$ known as primary Hopf surfaces. The complex structure moduli space is described by two parameters $p$, $q$. In the notation of~\cite{Assel:2014paa},  $p=e^{-2\pi |b_1|}$, $q = e^{-2\pi |b_2|}$ and $b_1$, $b_2$ are chosen real. The circle $S^1$ is parameterized by $\tau \sim \tau + 2\pi $. The three-sphere is described as a torus fibration over an interval: the torus angular coordinates are $\varphi_1 \sim \varphi_1 + 2\pi$ and $\varphi_2 \sim \varphi_2 + 2\pi$, while the interval coordinate is $\rho \in [0,1]$.
The metric is
\be
\label{4dmetrictoric}
\diff  s^2 \ = \ \Omega^2 \diff\tau^2 + \diff s^2(M_3) \ = \ \Omega^2 \diff\tau^2 + f^2 \diff\rho^2 +  m_{IJ}   \diff\varphi_I \diff\varphi_J \ ,
\ee
where $I,J=1,2$.
Supersymmetry imposes the constraint 
\be\label{OmegaToric}
\Omega^2 \ = \ b^I m_{IJ} b^J  \ ,
\ee
which ensures Hermiticity of the metric.
 $f$ and $m_{IJ}$ are arbitrary functions of $\rho$, except that $m_{IJ}$ must be positive definite and suitable boundary conditions need to be satisfied at the extrema of the interval. 
As $\rho \to 0$, we require that
\be\label{conditionsrhoto0}
f \to f_2 \, , \quad m_{11} \to m_{11}(0) \,,\quad m_{22}= (f_2\rho)^2 + \mathcal O( \rho^3)\,,\quad m_{12} = \mathcal O(\rho^2)~,
\ee
where $f_2>0$ and $m_{11}(0)>0$ are constants. Similar boundary conditions are taken at $\rho\to1$.  
The background field $A$ is given by
\be\label{realAforS1M3}
A \ =\  \frac{| b_1b_2|}{2\Omega^2f}\,\left (\Omega \sqrt{m} \right)' \left( \frac{\diff \varphi_1}{b_1} - \frac{\diff \varphi_2}{b_2} \right) + \frac{1}{2} \diff \omega\ ,
\ee
where $m = \det m_{IJ}$, a prime denotes derivative with respect to $\rho$, and
\be\label{choicebeta}
\omega \ =\ {\rm sgn}(b_1)\,\varphi_1+ {\rm sgn}(b_2)\,\varphi_2\,.
\ee
We will not need the expression of the remaining background field $V$.

The class of three-sphere metrics $\diff s^2(M_3)$ in~\eqref{4dmetrictoric} includes the elliptically squashed three-sphere $S^3_{\fb}$, first studied in the context of localization in \cite{Hama:2011ea}.
This is obtained by setting $b_1 = \tfrac{\beta \fb}{2\pi r_3} $, $b_2 =\frac{ \beta}{2\pi r_3\fb}$, redefining the coordinate $\rho$ into a coordinate $\vartheta \in [0,\pi/2]$ such that $f \diff\rho = [\fb^{-2}\sin^2\vartheta + \fb^{2}\cos^2\vartheta]^{1/2}\, \diff\vartheta$, and taking $m_{11} = \fb^{-2}\cos^2\vartheta$, $m_{22} = \fb^{2}\sin^2\vartheta$, $m_{12} = 0$. With these choices, $\Omega =( \tfrac{\beta}{2\pi r_3})^2$ and the expressions of the background fields also simplify. Everything that we will say applies to this particular case.

The complex vectors $K$ and $Y$ defined in \eqref{defvectors} take the form
\be
K \ = \ \frac{1}{2} \left[b_1 \frac{\partial}{\partial \varphi_1} + b_2 \frac{\partial}{\partial \varphi_2} - i \frac{\partial}{\partial\tau}\right]\,,
\label{torickv}
\ee
\be\label{exprY}
Y \ = \  -\frac{e^{i\omega}}{2}\left[\frac{1}{f}\frac{\partial}{\partial\rho} + i\, \frac{{\rm sgn}(b_1b_2)}{\Omega\sqrt{m}} \, b_I m_{IJ}\varepsilon^{JK}\frac{\partial}{\partial\varphi_K}\right] \ ,
\ee
where $\varepsilon^{JK}$ is the antisymmetric symbol, with $\varepsilon^{12}= +1$. In order to evaluate the differential operators \eqref{DiffOp} it will be useful to record that
\be\label{contrYA}
Y^\mu A_\mu \ = \  \frac{i}{4}\, e^{i\omega} \left[ \frac{1}{f}\left (\log(\Omega \sqrt{m}) \right)'   - \frac{{\rm sgn}(b_1b_2)}{\Omega\sqrt{ m}}   \, b_I m_{IJ}\varepsilon^{JK}{\rm sgn}(b_K) \right]\ ,
\ee
\be\label{contrKA}
K^\mu A_\mu \ = \  \frac{1}{2} K^\mu \partial_\mu \omega \ = \  \frac{|b_1| + |b_2|}{4}\ .
\ee

In principle, the action of the $4d$ chiral multiplet can be reduced to one dimension by expanding all the fields in an orthonormal basis of eigenfunctions of the relevant differential operators, and performing the integral over the three-manifold $M_3$. This yields a $1d$ action for infinitely many $1d$ multiplets. In practice, carrying out this reduction on a general background such as the one above is not feasible, as the eigenfunctions are not known. However, for the purpose of computing the Casimir energy there is no need to perform the complete reduction. Indeed, comparing the supersymmetry transformations~\eqref{ChiTransfoTwisted} with the one-dimensional ones in~\eqref{1dsusyTrans}, it is clear that the modes that do not satisfy the shortening conditions in \eqref{1stOrderConditionsB}, \eqref{1stOrderConditionsphi} are going to produce long multiplets in $1d$. As explained in section~\ref{SpectrumH}, these do not contribute to the Casimir energy. Therefore we can focus our attention on the subsector that does satisfy either one of the shortening conditions in \eqref{1stOrderConditionsB}, \eqref{1stOrderConditionsphi}.
Let us study the two cases in turn.

\subsubsection*{Reduction to 1d Fermi multiplets}

We start analyzing the conditions \eqref{1stOrderConditionsB}, which lead to $1d$ Fermi multiplets.
We Fourier expand the dependence of $B$ on the torus coordinates $\varphi_1$, $\varphi_2$ as
\be\label{expanB}
B(\rho,\varphi_1,\varphi_2,\tau) \ =\ \sum_{n_1,n_2} b_{n_1,n_2}(\tau)B_{n_1,n_2}(\rho)\, e^{-in_1\varphi_1 - in_2\varphi_2}\ ,
\ee
where $n_1$, $n_2$ are integer.
Recalling~\eqref{exprY}, \eqref{contrYA}
and that $B$ has $R$-charge $r-2$, one can see that the condition $\hat{\mathcal L}_Y B =  0$ yields for each choice of $n_1,n_2$:
\be
 \frac{1}{f}B'_{n_1,n_2} = \left[\frac{r-2}{2}\frac{\left (\log(\Omega \sqrt{m}) \right)' }{f}   -  \frac{{\rm sgn}(b_1b_2)}{\Omega\sqrt{m}} \, b_I m_{IJ}\varepsilon^{JK}\left(n_K  + \frac{r-2}{2}{\rm sgn}(b_K)\right)  \right]B_{n_1,n_2}\, .
\ee
This differential equation determines $B_{n_1,n_2}(\rho)$. The actual solution depends on the form of the metric functions and is not important. For our purposes it is sufficient to impose that the solution is normalizable, and thus find a restriction for the allowed values of $n_1$, $n_2$. To this end, it is sufficient to study the equation near the extrema of the interval parameterized by $\rho \in [0,1]$ and make sure that it does not develop singularities. Recalling the boundary conditions \eqref{conditionsrhoto0}, we find that near $\rho = 0$ the equation is solved by
\be
B_{n_1,n_2} \ = \ k\, \rho^{- n_2\,{\rm sgn} (b_2)} + \ldots  
\ee
for some constant $k$.
This is normalizable when $n_2\,{\rm sgn} (b_2) \leq 0$. A similar analysis of the behavior near $\rho = 1$ leads to require $n_1\,{\rm sgn} (b_1) \leq 0$. It is thus convenient to redefine
$n_1 \to -n_1 \,{\rm sgn} (b_1)$, $n_2 \to -n_2 \,{\rm sgn} (b_2)$. After this redefinition, the integers $n_1$, $n_2$ must take non-negative values, and the expansion~\eqref{expanB} becomes
\be
B\  =\ \sum_{n_1,n_2 \geq 0} b_{n_1,n_2}(\tau)B_{n_1,n_2}(\rho)\, e^{in_1 {\rm sgn} (b_1)\varphi_1 + in_2{\rm sgn} (b_2)\varphi_2}\ .
\ee

Recalling \eqref{torickv}, \eqref{contrKA}, we can also compute
\be
\hat{\scL}_K B \ = \ -\frac{i}{2}\sum_{n_1,n_2\geq 0} D_\tau b_{n_1,n_2}(\tau)B_{n_1,n_2}(\rho)\, e^{in_1 {\rm sgn} (b_1)\varphi_1 + in_2{\rm sgn} (b_2)\varphi_2}\ ,
\ee
with
\be\label{Dtaub}
D_\tau b_{n_1,n_2} \ = \ \left(\partial_\tau + \lambda^B_{n_1,n_2}\right) b_{n_1,n_2} \ ,
\ee
and
\be
\lambda^B_{n_1,n_2} \ = \  - |b_1| n_1 - |b_2|n_2 + \frac{r-2}{2}(|b_1| + |b_2|) \ .
\ee

We are now ready to perform the reduction to one dimension. Setting $\hat\scL_Y B = 0$ and $\phi = C= 0$, the chiral multiplet Lagrangian \eqref{ChiralLtwisted} becomes
\be\label{shortenedL1}
\mathcal L \ = \  \,2i \ti B \hat{\scL}_K B - \ti F  F \ ,
\ee
and the supersymmetry variations simplify to
\bea
 \delta B  &=&   F  \ , \qquad\qquad \quad\;\;\,\, \ti\delta B  \ = \  0\ , \nn\\ [2mm]
 \delta F  &=& 0 \ , \qquad \qquad\qquad \,\ti\delta F  \ =\  2i\, \hat{\mathcal L}_K B\ .
\eea
We can expand $F$ as
\be
F\  =\ \sum_{n_1,n_2 \geq 0} f_{n_1,n_2}(\tau)B_{n_1,n_2}(\rho)\, e^{in_1 {\rm sgn} (b_1)\varphi_1 + in_2{\rm sgn} (b_2)\varphi_2}\ 
\ee
(note that the dependence on $\rho,\varphi_1,\varphi_2$ is chosen the same as that of $B$, so $F$ also satisfies $\hat{\mathcal L}_Y F =0$). Similar expansions hold for $\ti B$ and $\ti F$. The action associated with the Lagrangian \eqref{shortenedL1} reduces to an action for an infinite set of Fermi multiplets in one dimension:
\be
S \ =\ \int \diff \tau\int_{M_3} \diff^3 x\, \sqrt{g} \,\mathcal L \ = \ \int \diff \tau\sum_{n_1,n_2\geq 0}\left(\ti b_{n_1,n_2}D_\tau b_{n_1,n_2} - \ti f_{n_1,n_2}  f_{n_1,n_2} \right)\ .
\ee
This is an infinite set of decoupled $1d$ multiplets, labeled by $n_1,n_2$. It is also straightforward to see that the $4d$ supersymmetry variations decompose into a set of $1d$  variations for the Fermi multiplets:
\bea
 \delta b_{n_1,n_2}  &=&   f_{n_1,n_2}  \ , \qquad\qquad \quad\, \ti\delta b_{n_1,n_2}  \ = \  0\ , \nn\\ [2mm]
 \delta f_{n_1,n_2}  &=& 0 \ , \qquad \qquad\qquad\quad\ti\delta f_{n_1,n_2}  \ =\  D_\tau b_{n_1,n_2}\ .
\eea
 From the form of the covariant derivative in~\eqref{Dtaub}, we see that $\lambda^B_{n_1,n_2}$ should be identified with the charge $\sigma$ in section~\ref{sec:SusyVacE}. More precisely, $\sigma_{\rm Fermi} = \frac{2\pi}{\beta}\lambda^B_{n_1,n_2}$.\footnote{In order to compare with section~\ref{sec:SusyVacE}, we must first rescale the $S^1$ coordinate as $\tau^{\rm new}=\frac{\beta}{2\pi}\tau^{\rm old}$, so that it has period $\beta$. Then we implement a Wick rotation $t=i\tau^{\rm new}$. Overall this gives $(\partial_\tau + \lambda) = i(\partial_t - \frac{2\pi i}{\beta} \lambda)$.} Recalling eq.~\eqref{contribHfermi}, the contribution of the infinitely many Fermi multiplets to the vacuum expectation values of $\Sigma$ and $H$ is thus
\be
\langle H_{\rm Fermi} \rangle \ =\ \langle \Sigma_{\rm Fermi} \rangle \ =\ \frac{\pi}{\beta}\sum_{n_1,n_2\geq 0} \lambda^B_{n_1,n_2} \ .
\ee 

\subsubsection*{Reduction to 1d chiral multiplets}

In the same way we can study the conditions~\eqref{1stOrderConditionsphi}, which define a reduction to $1d$ chiral multiplets.
With some foresight, we expand
\be
\phi(\rho,\varphi_1,\varphi_2,\tau) \ =\ \sum_{n_1,n_2} \phi_{n_1,n_2}(\tau)\Phi_{n_1,n_2}(\rho)\, e^{-in_1 {\rm sgn} (b_1)\varphi_1 - in_2{\rm sgn} (b_2)\varphi_2}\ .
\ee
Recalling~\eqref{exprY}, \eqref{contrYA}, the condition $\hat{\mathcal L}_{\overline Y}\, \phi  = 0$ yields for each choice of $n_1,n_2$ :
\be
\frac{1}{f}\Phi'_{n_1,n_2}  =  \left[-\frac{r}{2}\frac{\left (\log(\Omega \sqrt{m}) \right)' }{f}  +  \frac{{\rm sgn}(b_1b_2)}{\Omega\sqrt{m}} b_I m_{IJ}\varepsilon^{JK}\left(n_K {\rm sgn}(b_K)  + \frac{r}{2}{\rm sgn}(b_K)\right)  \right]\Phi_{n_1,n_2}\, .
\ee
This determines $\Phi_{n_1,n_2}(\rho)$. By studying the equation near $\rho = 0 $ and $\rho = 1$, we see that normalizability requires $n_1\geq 0$, $n_2\geq 0$.
We also compute
\be
\hat{\mathcal L}_K  \phi \ = \ -\frac{i}{2}\sum_{n_1,n_2 \geq 0} D_\tau \phi_{n_1,n_2}(\tau)\Phi_{n_1,n_2}(\rho)\, e^{-in_1 {\rm sgn} (b_1)\varphi_1 - in_2{\rm sgn} (b_2)\varphi_2} \ ,
\ee
with
\be\label{D_tau_phi}
D_\tau \phi_{n_1,n_2} \ = \ \left(\partial_\tau + \lambda^\phi_{n_1,n_2}\right) \phi_{n_1,n_2} \ ,
\ee
and
\be
\lambda^\phi_{n_1,n_2} \ =\  |b_1|n_1 + |b_2|n_2 + \frac{r}{2}(|b_1| + |b_2|) \ .
\ee

Setting $\hat{\scL}_{\overline Y} \phi = \hat{\mathcal L}_{\overline Y}C = B = F =0$, the $4d$ Lagrangian~\eqref{ChiralLtwisted} becomes
\be\label{shortenedL2}
\mathcal L \ =\  4\, \scL_{\overline K} \ti \phi\, \hat{\scL}_K \phi  + i \kappa\big( \hat{\scL}_K \ti \phi\, \phi - \ti \phi \hat{\scL}_K \phi \big) + 2 i\,\ti C \hat{\scL}_{\overline K}C - \kappa \ti C C\ , 
\ee
with supersymmetry variations
\bea
 \delta \phi &=&   C \ , \qquad \qquad\qquad  \ti\delta \phi \ =\  0 \ ,\nn\\ [2mm]
 \delta C  &=&  0 \ , \qquad\qquad\qquad\, \ti\delta C  \ =\  2i\, \hat{\mathcal L}_K \phi\ .
 \eea
By expanding $C$ as
\be
C\  =\ \sum_{n_1,n_2 \geq 0} c_{n_1,n_2}(\tau)\Phi_{n_1,n_2}(\rho)\, e^{-in_1 {\rm sgn} (b_1)\varphi_1 - in_2{\rm sgn} (b_2)\varphi_2}\ ,
\ee
performing similar expansions for $\ti\phi$, $\ti C$
and integrating over the four-manifold, we obtain an infinite set of decoupled one-dimensional chiral multiplets.
 However, the essential information that we need is more straightforwardly extracted from the supersymmetry transformations. For any choice of $n_1,n_2$, these read
\bea
 \delta \phi_{n_1,n_2} &=&   c_{n_1,n_2} \ , \qquad \qquad\qquad  \ti\delta \phi_{n_1,n_2} \ =\  0 \ ,\nn\\ [2mm]
 \delta c_{n_1,n_2}  &=&  0 \ , \qquad\qquad\qquad\qquad \ti\delta c_{n_1,n_2}  \ =\  D_\tau \phi_{n_1,n_2}\ .
 \eea
By comparing with the supersymmetry variations given in \eqref{1dsusyTrans}, we see that $\lambda^\phi_{n_1,n_2}$ is proportional to the charge $\sigma$ of the $1d$ chiral multiplet: $\sigma_{\rm chiral} = \frac{2\pi}{\beta}\lambda^\phi_{n_1,n_2}$. Then using~\eqref{contribHchiral} the contribution of the $1d$ chiral multiplets to the vacuum expectation value of $\Sigma$ and $H$ is
\be
\langle H_{\rm chiral} \rangle \ =\ \langle \Sigma_{\rm chiral} \rangle \ =\ \frac{\pi}{\beta}\sum_{n_1,n_2 \geq 0} \lambda^\phi_{n_1,n_2} \ .
\ee

\subsubsection*{The vacuum energy}

The VEV of the full one-dimensional Hamiltonian is given by
\be\label{infinitesum}
\langle H_{\rm susy}\rangle \ = \ \frac{\pi}{\beta}\sum_{n_1,n_2 \geq 0} \lambda^\phi_{n_1,n_2} + \frac{\pi}{\beta}\sum_{n_1,n_2\geq 0} \lambda^B_{n_1,n_2} \ ,
\ee
with 
\bea
\lambda^\phi_{n_1,n_2} &=&  |b_1|n_1 + |b_2|n_2 + \frac{r}{2}(|b_1| + |b_2|)  \ ,\nn \\ [2mm]
\lambda^B_{n_1,n_2} &=&  - |b_1| n_1 - |b_2|n_2 + \frac{r-2}{2}(|b_1| + |b_2|) \ .
\eea
These infinite sums are divergent and require regularization. As in subsection~\ref{ReductionRoundS3}, we will regularize the two sums separately, using the Barnes double zeta function. For simplicity in the next expressions we assume $b_1>0$, $b_2>0$; the formulae however hold more generally, with $b_1$, $b_2$ replaced by $|b_1|$, $|b_2|$.

The sum \eqref{infinitesum} can be written as
\be\label{DIFFzeta2}
\langle H_{\rm susy} \rangle \ = \  \frac{1}{2}\,\zeta_2\left(-1;b_1,b_2, \tfrac{r}{2}(b_1+b_2)\right) - \frac{1}{2}\,\zeta_2\left(-1;b_1,b_2, \tfrac{2-r}{2}(b_1+b_2)\right) \ ,
\ee
where $\zeta_2$ is the Barnes double zeta function, defined as
\be
\zeta_2(s;b_1,b_2,x) \ = \ \sum_{n_1,n_2 \geq 0} (b_1n_1+b_2n_2 +x)^{-s}\ ,
\ee
with $b_1$, $b_2$, $x$ real and positive.
At $s = -1$, it evaluates to (see {\it e.g.}~\cite{Spreafico})
\be
\zeta_2(-1;b_1,b_2, x) \; =\; - \frac{b_1+b_2}{24} + \left(3+\frac{b_1}{b_2} + \frac{b_2}{b_1} \right) \frac{x}{12} - \left(\frac{1}{b_1} + \frac{1}{b_2}\right) \frac{x^2}{4} + \frac{x^3}{6b_1b_2}  \ .
\ee
This gives for the vacuum energy
\be
\langle H_{\rm susy} \rangle \ = \ \frac{4\pi}{3\beta}\left(b_1 + b_2\right)(a-c) + \frac{4\pi}{27\beta}\frac{(b_1+b_2)^3}{b_1b_2}(3c-2 a) \ .
\ee
This is the same expression appearing in eq.\ (5.10) of~\cite{Assel:2014paa}. Redefining  $b_1 = \frac{\beta \fb}{ 2\pi r_3}$ and $b_2 = \frac{\beta \fb^{-1}}{ 2\pi r_3 }$, where $\beta$ coincides with the length of $S^1$, gives the result \eqref{squashedC} advertised in the introduction.

\subsubsection*{Back to the round case}

Let us conclude by coming back to the $S^3 \times S^1$ background with $SU(2)_l \times SU(2)_r \times U(1)$ symmetry discussed in subsection~\ref{ReductionRoundS3}. The round sphere is just $S^3_\fb$ with $\fb=1$. In this case $|b_1|=|b_2|=\frac{\beta}{2\pi r_3}$ and the result above for the Casimir energy simplifies to the result in subsection~\ref{ReductionRoundS3}. 
In more detail, the sum~\eqref{infinitesum} becomes
\bea
\langle H_{\rm susy}\rangle &=& \frac{1}{2r_3}\sum_{n_1,n_2 \geq 0} \left(n_1 + n_2 + r\right) + \frac{1}{2r_3}\sum_{n_1,n_2\geq 0} \left(-n_1-n_2 + r-2\right)\nn \\ [2mm]
&=& \frac{1}{2r_3}\sum_{\ell \geq 0} \left(\ell + 1\right) \left(\ell + r\right) - \frac{1}{2r_3}\sum_{\ell\geq 0} \left(\ell + 1\right)\left( \ell + 2 -r\right) \ ,
\eea
where in the second line we have defined $\ell = n_1+n_2$. So the regularization used above and the one used in subsection~\ref{ReductionRoundS3} are compatible.

It is also straightforward to see that the shortening conditions match. When expanding the chiral multiplet $(\phi,C,B,F)$ in scalar spherical harmonics, the differential operators discussed above become
\be
 \hat{\mathcal L}_Y  \ = \ i L_{l}^- \ , \qquad \qquad \hat{\mathcal L}_K \ = \  -\frac{i}{2} \left( L^3_l + \nabla_\tau +  \frac{q_R}{2} \right) \ ,
\ee
where $q_R$ is the charge of the field the operator is acting on, and we introduced the $SU(2)_l$ generators $L_l^{\pm}, L_3$.
 Then the condition $\hat{\mathcal L}_{\overline Y} \phi  = 0$ is nothing but $L_l^+ Y_{\ell, m,n} = 0$, which selects the harmonics with highest quantum number $m$, namely $Y_{\ell, \frac{\ell}{2},n}$. The condition $\hat{\mathcal L}_Y B = 0$ translates into $L_l^- Y_{\ell, m,n} = 0$ and thus selects $Y_{\ell, -\frac{\ell}{2},n}$.

\subsection*{Acknowledgements}
We would like to thank H.-C.~Kim, S.~Kim, S.~Minwalla, and A.~Schwimmer for useful discussions.  B.A., J.L., and D.M. acknowledge support by the ERC Starting Grant N. 304806, ``The Gauge/Gravity Duality and Geometry in String Theory.'' D.C. is supported by an European Commission Marie Curie Fellowship under the contract PIEF-GA-2013-627243. Z.K. and L.D.P are supported by the ERC STG grant 335182, by the Israel Science Foundation under grant 884/11, by the United States-Israel Binational Science Foundation (BSF) under grant 2010/629, as well as by the I-CORE Program of the Planning and Budgeting Committee and by the Israel Science Foundation center for excellence grant (grant no. 1989/14). Any opinions, findings, and conclusions or recommendations expressed in this material are those of the authors and do not necessarily reflect the views of the funding agencies.
 


\appendix


\section{Casimir energy in CFT in $d=2$ and $d=4$}
\label{casimappendix}

In this Appendix we consider a CFT, not necessarily supersymmetric, in $d$ dimensions coupled to a background metric $g_{\mu\nu}$ with Euclidean signature. 
We denote the generating functional of connected correlation functions by $\scW[g] = - \log Z[g]$. Derivatives with respect to the metric give insertions of the energy-momentum tensor
\bea 
\label{emtensor}  \langle \sqrt{g} \, T^{\mu\nu} \rangle &  =  &  2{\delta \scW \over \delta g_{\mu\nu}} ~.
\eea

Under an infinitesimal Weyl transformations $g_{\mu\nu} \to (1+2\sigma) g_{\mu\nu}$, the transformation of $\scW$ is given by the Weyl anomaly. In $d=2$ and $d=4$ respectively we get 
\be\label{anomaly2d}  \delta_\sigma \scW \ = \ \int \diff^2 x\sqrt{g}\, \sigma \left(-{c \over 24\pi} R\right)~,\ee
\be\label{anomaly}  \delta_\sigma \scW \ = \ \frac{1}{(4\pi)^2} \int \diff^4 x\sqrt{g}\, \sigma \left( a E_{(4)} - c W^2\right)~.\ee
We have set the coefficient $b$ that appears in~\eqref{localterm} to zero. The above infinitesimal Weyl transformations can be integrated to the so-called Dilaton action,\footnote{We thank Adam Schwimmer for discussions.} see {\it e.g.}~\cite{Fradkin:1983tg,Brown:1985ri,Cappelli:1988vw}: 
\be\label{diffW} \scW[e^{2\sigma}g] - \scW[g] \ = \   - S_{\rm D}[-\sigma, g] \ = \  S_{\rm D}[\sigma, e^{2\sigma} g]~, \ee
where $S_{\rm D}$ has the following explicit expressions in $d=2$ and $d=4$
\begin{align} S^{d=2}_{\rm D}[\sigma,g_{\mu\nu}] & \ = \  {c\over 24\pi} \int \diff^2x\sqrt {g}  \left(- \sigma R+(\partial_\mu\sigma)^2\right) ~,  \end{align}
\begin{align} S^{d=4}_{\rm D}[\sigma,g_{\mu\nu}] & = \frac{a}{(4\pi)^2} \int \diff^4x\sqrt {g}  \left(\sigma E_4+4\partial_\mu\sigma\partial_\nu\sigma\left(R^{\mu\nu}-\tfrac12 g^{\mu\nu}R\right)-4(\partial\sigma)^2\square\sigma+2(\partial\sigma)^4 \right)  \nonumber \\ & - \frac{c}{(4\pi)^2} \int \diff^4x\sqrt {g} \, \sigma\, W^2 ~.  \end{align}
Taking a derivative of \eqref{diffW} with respect to the metric, we find
\be\label{vevdiff}
e^{d\sigma}\langle \, T^\mu_{~\nu} \rangle_{e^{2\sigma} g} - \langle  \, T^\mu_{~\nu} \rangle_{g} \ = \  \frac{2}{\sqrt{g}} g_{\nu\lambda}\frac{\delta( - S_{\rm D}) }{ \delta g_{\mu\lambda}}[-\sigma, g]~.
\ee
We see that the change in the VEV of the energy-momentum tensor under Weyl rescaling is fixed by the Dilaton action.

\subsection{The cylinder and its infinitesimal deformation}

In the case of a conformally flat geometry with metric $e^{2\sigma}\delta_{\mu\nu}$ (such as $S^{d-1}\times \R$) we can use the fact that the VEV in flat space is unambiguously fixed to vanish, {\it i.e.}\ $\langle T^\mu_{~\nu} \rangle_{\R^d} = 0$. One then finds for the energy-momentum tensor on this space
\be  \langle T^\nu_{~\mu}  \rangle^{(d=2)}_{e^{2\sigma}\delta}  \ = \  {c\over 12 \pi} e^{-2\sigma}  \left(\Box \sigma  \delta^\nu_{~\mu} - \partial^\nu \partial_\mu \sigma + \partial^\nu \sigma \partial_\mu \sigma - \half \delta^\nu_{~\mu} (\partial \sigma)^2\right)~, \ee  
\be  
\langle T^\nu_{~\mu}  \rangle^{(d=4)}_{e^{2\sigma}\delta}  \ = \   - \frac{a}{(4\pi)^2}  \left.  \left[ \delta^\nu_{~\mu} \left(R^{\rho\tau}R_{\rho\tau} - \half R^2\right) - 2 R^{\nu\rho}R_{\rho\mu} +{4 \over 3} R R^\nu_{~\mu} \right] \right|_{e^{2\sigma}\delta }~. 
\ee  
These results were first derived in \cite{Brown:1977sj} using a different method. 

Consider in particular the round cylinder $S^{d-1} \times \R$ with the conformally flat metric
\begin{align}\label{roundmetric}
\diff s^2_{S^{d-1} \times \R} & \ = \   \diff\tau^2 + r_{d-1}^2 \diff \Omega_{d-1}^2 \ = \   \left(\frac{r_{d-1}}{r} \right)^2(\diff r^2 + r^2 \diff\Omega_{d-1}^2)~.
\end{align}
Here $\diff\Omega_{d-1}^2$ denotes the metric on a $(d-1)$-sphere with unit radius, and $r_{d-1}$ is the radius of the sphere. Evaluating the energy-momentum tensor in this case, one finds the following non-zero components of $\langle T^\mu_{~\nu} \rangle$ in $d=2$:
\be
T^\tau_{~\tau} \  = \ -{c\over 24\pi r_1^2} ~, \qquad\qquad T^\theta_{~\theta} \ =\ {c\over 24\pi r_1^2}~,
\ee
and in $d=4$:
\be\label{VEVround}
T^\tau_{~\tau} \ = \ \frac{3a}{8\pi^2r_3^4}~,\qquad\qquad
T^i_{~j} \ =\ -\frac{a}{8\pi^2r_3^4}\,\delta^i_{~j}~.
\ee
Note that $\langle T^\mu_{~\mu} \rangle_{S^3\times \R} = 0$, because the Weyl anomalies evaluate to zero on the cylinder. We can easily reintroduce the $b$ dependence of the result in $d=4$ by taking a functional derivative of the associated counterterm $-\frac{b}{12(4\pi)^2} \int \diff^4x \sqrt{g} R^2$. This results in a shift of the coefficient $a \to a - \frac{b}{2}$.

Integrating $\langle T^\tau_{~\tau} \rangle$ over $S^1$ and $S^3$ respectively gives the Casimir energies quoted in the text~\eqref{C2d} and~\eqref{nobox}.

Above we have reviewed how to compute the Casimir energy on a conformally flat background $e^{2\sigma}\delta_{\mu\nu}$. We will now slightly extend our considerations by taking a geometry that is a generic small perturbation of a conformally flat one, {\it i.e.}\ $g_{\mu\nu} = e^{2\sigma}\delta_{\mu\nu} + h_{\mu\nu} $. Let us compute the change in the ground state energy to first order in $h_{\mu\nu}$. The idea is that we can infer the ground state energy from the change in the partition function $\scW$ as a result of the perturbation. 

This can be  approached as follows: from the definition \eqref{emtensor} we have 
\be
\delta \scW\ =\ \half \int \diff^4x\sqrt {g}\,\langle T_{\mu\nu}\rangle_{e^{2\sigma} \delta}\, h^{\mu\nu}+\scO(h^2)~,
\label{deltaW}
\ee
so we just need to know $\langle T_{\mu\nu}\rangle_{e^{2\sigma}\delta}$, {\it i.e.}\ the VEV in the conformally flat original space. 

For instance, consider the round cylinder $ S^3\times \R$ with a time-independent perturbation $h_{ij}$ of the metric on the three-sphere. The result for the VEV of the energy-momentum tensor on the cylinder was given in equation \eqref{VEVround}. Therefore we find that 
\be
\delta \scW\ =\ - \frac{ (2 a - b)}{32\pi^2 r_3^4} \int \diff^4x\sqrt{g}\, h^i_{~i}~.
\ee
Since the metric perturbation is assumed time-independent, we can interpret this integral as \be\label{energy}-\frac{(2 a-b)}{32\pi^2 r_3^4} \int \diff^4x\sqrt{g}\, h^i_{~i}\ =\ - \frac{ (2 a - b)}{32\pi^2 r_3^4} \int \diff^3x\sqrt{g}\, h^i_{~i} \int  \diff\tau\ee and the coefficient of $\int \diff\tau$ gives the correction to the energy of the ground state. The final result for the Casimir energy is
\be\label{correctionE} E_0\ =\ (2a - b)\left(\frac{3} {8r_3} -\frac{1}{32\pi^2 r_3^4} \int \diff^3x\sqrt{g}\, h^i_{~i} \right)+\scO{(h^2)}~.  \ee
We see that to order $\scO(h)$ the Casimir energy is still scheme dependent, being proportional to the same combination $2a-b$ as the leading order result.
However, the ratio of the leading term and the subleading term is unambiguous (in fact, the ratio is fixed by imagining a small perturbation that only changes the radius of the three-sphere).

One can speculate that going to second order in the perturbation $\scO{(h^2)}$ the $c$-anomaly will also appear and that some terms will be scheme independent. Notice, however, that at this order a new counterterm, proportional to $\int \diff^4x\sqrt g\,W^2$, could contribute. 

\subsection{The Casimir energy and holography}

We now summarize the status of the Casimir energy in  the context of the gauge/gravity duality.  We will mainly focus on four-dimensional CFTs admitting a
dual description in terms of solutions of type IIB supergravity of the type AdS$_5\times M_5$, and their deformations. 
Similar considerations can be made for six-dimensional CFTs with AdS$_7$ gravity duals.  

The comparison of the Casimir energy of a four-dimensional CFT on $S^3 \times \mathbb R$ with the expectation value of the holographic  energy-momentum tensor was first made in 
\cite{Balasubramanian:1999re}, although in this reference the authors focussed on ${\cal N}=4$ SYM and its AdS$_5\times S^5$ dual, and some of the comments
 made there apply only to this example. Let us review the discussion of \cite{Balasubramanian:1999re}.  
Using the holographic energy-momentum tensor, computed in global  AdS$_5$, one obtains for the Casimir energy, defined exactly as in (\ref{cas}), the expression\footnote{This was
referred to as ``mass of global AdS$_5$'' in \cite{Balasubramanian:1999re}. We have  used the standard formula
for the Newton constant $ G_5^{-1}=\frac{2N^2}{\ell^3}\pi^4 \mathrm{vol}' (M_5)$, where $\ell$ is the AdS radius and in the specific case vol$'(S^5)=\pi^3$.}
\bea
E_0^\mathrm{holo} & = & \frac{3N^2}{16\ell}~.
\label{holoe0}
\eea
It is important to note that this result is obtained using a minimal holographic renormalization scheme, where there are no finite counterterms $\Delta S_{\rm ct}$ added to the on-shell action. 
This  corresponds to the absence of the $\Box R$ term in the holographic trace anomaly $\langle T^{\mu}_\mu\rangle$. In other words, it corresponds to the scheme $b=0$.

The expression (\ref{holoe0}) is compared with a Casimir energy in a free CFT, comprising $n_0$ real scalars, $n_{1/2}$ Weyl fermions, and $n_1$ Abelian gauge fields. 
In particular, the scalar fields considered are conformally coupled, and the fermions are massless. This results in the expression 
\bea
E_0^{\rm free} & = & \frac{1}{960 r_3}(4n_0 + 17 n_{1/2} + 88 n_1)~,
\label{freee0}
\eea
obtained summing up the contributions of the single fields, which in turn are regularised using zeta function. For the specific case of the ${\cal N}=4$ SYM theory, we have 
\bea
E_0^{\rm free} & = & \frac{3(N^2-1)}{16r_3}  \  = \ \frac{3}{4r_3}a \qquad \mathrm{for}~~{\cal N}=4 ~~\mathrm{SYM}~,
\eea
which agrees with (\ref{holoe0}) at leading order in $N$ 
\cite{Balasubramanian:1999re}. 

However, the agreement of $E_0^\mathrm{holo}$ with $E_0^{\rm free}$ is, in some sense, accidental, and may be misleading. Indeed for generic $n_0$, $n_{1/2}$, $n_1$, the expression~\eqref{freee0} for $E_0^{\rm free}$ \emph{is not} proportional to the $a$ anomaly: it also includes a contribution from $b$. In fact, the scheme leading to \eqref{freee0} generically also gives a $\square R$ contribution to the trace anomaly. Once the particular matter content of ${\cal N}=4$ SYM is specified, the coefficient $b$ happens to vanish (see {\it e.g.}~\cite{Fradkin:1983tg}) and the scheme leading to~\eqref{freee0} effectively coincides with the scheme $b=0$ which was used in the holographic computation. 

More generally, one can ask whether the computation of the Casimir energy in a deformed cylinder can be reproduced by a five-dimensional supergravity solution deforming AdS$_5$.  
At least at leading order in the deformation, there exists a holographic counterpart to the formula  (\ref{deltaW}). Consider a one-parameter family of deformations of the \emph{boundary} metric, with parameter $\mu$. 
Then by simply applying the chain rule to the renormalised on-shell action one obtains\footnote{In  \cite{Cassani:2014zwa} this identity was written to include the variation of the boundary gauge field. 
However, if we are not interested in supersymmetry, we can vary the metric independently.} 
\bea
\frac{\diff }{\diff \mu} S_{\rm ren} &=& \int_{\partial M_5} \diff^4 x \sqrt{g} \left( -\frac{1}{2} \langle T_{\mu\nu}\rangle \frac{\diff g^{\mu\nu}}{\diff \mu}   \right)~,
\label{WardIdentityGeneral}
\eea
where $g_{\mu\nu}$ is the finite metric on the boundary, and  $\langle T_{\mu\nu}\rangle$ is the holographic energy-momentum tensor. Setting $g=g_{(0)} + \mu h$, with $\mu$ infinitesimal, and expanding 
(\ref{WardIdentityGeneral}) at first order in $\mu$ leads exactly to the holographic version of the formula  (\ref{deltaW}). Thus, the holographic 
Casimir energy on an  infinitesimally squashed cylinder is guaranteed to agree with the field theory result. In the next subsection, we verify this in an explicit example.

For  the supersymmetric version of the Casimir energy the situation is different.  One of the main points emphasized in this paper is that on the field theory side, 
computing in the free limit gives a reliable result for $E_{\rm susy}$,  valid also at strongly coupled  points. However, how to reproduce this in the gravity side remains an open 
problem. In  \cite{Cassani:2014zwa} it was suggested that a supersymmetric treatment of holographic renormalization might reveal the existence of new boundary terms, 
that would lead to a matching of the on-shell action of existing solutions with $E_{\rm susy}$. Another possibility is that there exist other (Euclidean) supersymmetric 
solutions, whose on-shell action would match with  $E_{\rm susy}$, using the standard holographic renormalization technology. In light of the results of the present paper, 
it will be very interesting to revisit this problem. 

Let us also mention that there exists a similar open problem in the context of SCFTs on the six-dimensional cylinder $S^5\times \mathbb R$. A supersymmetric version of the Casimir energy was discussed in 
\cite{Kim:2012ava}, where it was also noted that it does not match the standard holographic Casimir energy in AdS$_7$. It  may be useful to prove that this quantity is physical in 6d SCFTs by
performing an analysis like the one in the present paper,  and to investigate the seven-dimensional holographic dual.

\subsection{Holographic check of $E_0$ on a squashed cylinder}

Below we will compare the result of the first-order correction to the ordinary Casimir energy due to a non-conformally flat geometry~\eqref{correctionE}, with a corresponding 
holographic result, that can be easily extracted from the gravity solution presented in \cite{Cassani:2014zwa}. 
Although the solution in \cite{Cassani:2014zwa} is supersymmetric, we can  obtain from this an expression for the Casimir energy, 
that is valid independently of supersymmetry. 
In order to make the discussion as  self-contained as possible, we will begin recalling  relevant aspects of the solution in \cite{Cassani:2014zwa}, referring the reader to this reference for more details. 

This is a one-parameter family of supersymmetric solutions deforming AdS$_5$, constructed as an asymptotically locally anti de Sitter (AlAdS)
solution of five-dimensional minimal gauged supergravity. 

An analytic continuation to Euclidean signature yields the gravity dual to a class of four-dimensional ${\cal N}=1$ 
supersymmetric gauge theories on a curved manifold with topology $S^3 \times S^1$. In particular, the boundary metric is that of a  squashed three-sphere
preserving  $SU(2)\times U(1)$ isometry, and there is a non-trivial  background gauge field coupling to the $R$-symmetry current.  So, this is an instance of a  Hopf surface, as discussed in the main part of the text \cite{Cassani:2014zwa}.

In a coordinate system, the five-dimensional metric and the graviphoton field take the following asymptotic form near to the boundary, corresponding to 
$\rho \to \infty$:
\bea
\label{asymptoticmetric}
\diff s^2_{5d} & = & \diff \trho^2 + e^{2\trho/\ell} \, \diff s^2_{\rm bdry}\,+\, \ldots\,, \nonumber\\
\label{asymptoticA}
A_{5d} & = & A_{\rm bdry} \,+\, \mathcal O(e^{-\trho})\,,
\eea
where the boundary values read 
\be\label{bdrymetric}
\diff s^2_{\rm bdry} \ = \  (2a_0)^2\left[ - \frac{1}{\sq^2} \diff t^2 + \frac{\ell^2}{4}\left(\sigma_1^{\,2} + \sigma_2^{\,2}  +   \sq^2\sigma_3^{\,2}\right) \right]~,
\ee
and
\be\label{bdryA}
A_{\rm bdry}\ =\  
\frac{1}{2\sqrt 3} \left[ \frac{\diff t}{\ell} + (\sq^2-1) \sigma_3\right]\, ,
\ee
respectively. Here $\ell$ is the AdS radius, that can be identified with the radius of the $S^3$, and the left-invariant one-forms  $\sigma_a$'s are defined as usual 
(see \cite{Cassani:2014zwa}). The full five-dimensional solution is determined in terms of the single  parameter $v$, measuring the squashing of the boundary metric. 
For $v^2=1$ the solution reduces to AdS$_5$ and the boundary metric is conformally flat. The parameter $2a_0$ is an overall scale, 
that  can be set to any value by simply shifting the radial coordinate $\rho$.\footnote{In~\cite{Cassani:2014zwa} the parameter $a_0$ is chosen to be a convenient function of $v$ (so that the solution ends at $\rho =0$). However, this is irrelevant for the present discussion.} Notice that the background gauge field in (\ref{bdryA}) comprises a constant part along $\diff t $, that is necessary to have well defined supercharges in the compactified geometry, as discussed around~ (\ref{supalii}).  However, in order to compare with the result above, which is valid in the \emph{absence} of such 
term\footnote{This term maps to a gauge field $A\sim \tfrac{\diff r}{r}$ that is  \emph{singular} at the origin of $\mathbb{R}^4$; this is why we do not include it in our treatment.} (and independently of supersymmetry), we should remove this by shifting the graviphoton field as 
 $A_{5d} \to A_{5d} -  \frac{1}{2\sqrt 3\ell}  \diff t$. It is simple to check that the solution with this new gauge field is still non-singular, and its relevant properties may be extracted from 
 \cite{Cassani:2014zwa}, by following through this simple change of gauge.  
 
 After Wick rotating to Euclidean signature, and setting $v=1+\epsilon$, we have at leading order the following boundary  metric
 \be
\diff s^2_{\rm bdry}  =\diff s^2_{(0)} + \diff s^2_{(1)} + \mathcal O(\epsilon^2) \ ,
\ee
with
\bea\label{order0order1}
\diff s^2_{(0)} & = &  g^{(0)}_{\mu\nu}\diff x^\mu \diff x^\nu\ =\   \diff \tau^2 + \frac{\ell^2}{4}\left( \sigma_1^2 + \sigma_2^2 + \sigma_3^2 \right) ~, \nonumber\\
\diff s^2_{(1)} & = & h_{\mu\nu}\diff x^\mu \diff x^\nu\ = \  \epsilon \left[ - 2 \diff \tau^2 + \frac{\ell^2}{2} \sigma_3^2 \right]\ ,
\eea
and background gauge field 
\be
A_{\rm bdry}\ =\  
\frac{\epsilon}{\sqrt 3} \, \sigma_3   + {\cal O} (\epsilon^2)\,.
\ee
In particular, this implies that the contribution of the background $R$-symmetry current to the Casimir energy through a term of the type 
$\int_{S^3}\diff^3 x\sqrt{g} \,\langle  J_R^\mu \rangle  A_\mu$  can only affect the result at order ${\cal O}(\epsilon^2)$. Therefore at linearized order we can neglect the background gauge field.  

Using \eqref{VEVround}, \eqref{deltaW} we can evaluate the correction to the Casimir energy due to the perturbation in~\eqref{order0order1}. We obtain
\be
\delta E_0 \ = \ \frac{1}{2}(2\pi^2\ell^3)\left(-\epsilon\,\frac{ a}{\pi^2\ell^4}\right) \ = \  - \epsilon\,\frac{a }{\ell}~,
\ee
where the factor $ 2\pi^2\ell^3$ comes from
the integration over the three-sphere.
Thus, the Casimir energy at first order in the squashing reads
\be
E_0  \ = \ \left(\frac{3}{4} - \epsilon\right)\frac{a}{\ell} \ .
\label{smallecft}
\ee

This can be compared with the result of the holographic computation. The Euclidean on-shell action of five-dimensional supergravity was computed exactly 
as a function of the parameter $v$ in \cite{Cassani:2014zwa}, and reads
\be\label{OnShAct}
S \ = \ \frac{8a}{\ell} \left( \frac{2}{27v^2} + \frac{2}{27} -\frac{13}{108}v^2 + \frac{19}{288}v^4\right) \int \diff\tau\ ,
\ee
where we rewrote the five-dimensional Newton constant in terms of the $a$ anomaly of the dual CFT. 
This was obtained in a gauge such that $A^{\rm bdry}_\tau \neq 0$, so we should shift the gauge to $A^{\rm bdry}_\tau =0$ and correspondingly the on-shell action using eq.~(4.14) therein. In fact, the shift due to the change of gauge is proportional to $\epsilon^2$, so at linear order, this shift is immaterial. In any case, after doing this shift,  we obtain exactly the expression appearing in (4.35) therein for $E_0$, times $\int \diff \tau $. 
The on-shell action (\ref{OnShAct}) expanded at first order in $\epsilon$, gives
\be
S \ =\ \left(\frac{3}{4} - \epsilon\right)\frac{a}{\ell} \int \diff \tau \ ,
\ee
thus we get  perfect agreement with the dual field theory result (\ref{smallecft}).
Note that (\ref{OnShAct}) has been evaluated using a holographic renormalization 
scheme without  the $\int \diff^4 x \sqrt{g} R^2$ counterterm, precisely as in the conformally flat background.


\section{$E_{\rm susy}$ and the Hamiltonian}
\label{app:EvsH}

In this Appendix we show that the supersymmetric vacuum energy $E_{\rm susy}$, defined within the path integral approach as 
\be
E_{\rm susy} \ = \ -\lim_{\beta\to \infty}\frac{{\rm d}}{{\rm d}\beta}\log Z^{\rm susy}_{M_3 \times S^1_\beta}\ ,
\ee
coincides with the VEV of the charge associated to the time translation symmetry. In other words, $E_{\rm susy}$ is the VEV of the Hamiltonian, $E_{\rm susy}  =  \left\langle H_{\rm susy}  \right\rangle\,$. This is well-known,  in particular in the context  of quantum field theories at non-zero temperature. Nevertheless in the presence of additional non-dynamical background fields it may be useful to spell   out some details.

$E_{\rm susy}$ can be expressed in terms of the energy-momentum tensor and the currents appearing in the $R$-multiplet in the following way. Consider a background $M_3 \times S^1_\beta$ as in the main text, then perform a trivial rescaling of the $S^1$ coordinate $\tau$ so that $\tau \sim \tau + 1$ and the background fields ${g}_{\tau\tau}$, $A_\tau$ and $V_\tau$ acquire a dependence on $\beta$. Applying the chain rule to the variation of $-\log Z$ with respect to $\beta$, $E_{\rm susy}$ can be written as
\bea\label{Esusy_bckgd_fields}
E_{\rm susy} &=&  \lim_{\beta\to \infty}\left\langle \int\diff^4 x \sqrt g \left(-\frac 12  T_{\mu\nu} \,\frac{{\rm d}g^{\mu\nu}}{{\rm d}\beta} +  J_{\rm R}^\mu \, \frac{{\rm d} A_\mu}{{\rm d}\beta} -\frac 32  J_{\rm FZ}^\mu \, \frac{{\rm d} V_\mu}{{\rm d}\beta}\right)\right\rangle \nn \\ [2mm]
&=& \lim_{\beta\to \infty}\frac{1}{\beta} \left\langle \int\diff^4 x \sqrt g \left(  T^\tau{}_{\tau}   +  J_{\rm R}^\tau \,  A_\tau -\frac 32  J_{\rm FZ}^\tau \,  V_\tau \right) \right\rangle\ ,
\eea
where the energy-momentum tensor $T_{\mu\nu}$, the $R$-current $J^\mu_{\rm R}$ and the Ferrara-Zumino current $J^\mu_{\rm FZ}$ are defined as
\be\label{defT_JR_JFZ}
T_{\mu\nu} = -\frac{2}{\sqrt{g}}\frac{\delta S}{\delta g^{\mu\nu}}\ , \qquad J^\mu_{\rm R} \ = \ \frac{1}{\sqrt g} \frac{\delta S}{\delta A_\mu} \ , \qquad - \frac 32 J^\mu_{\rm FZ} \ = \ \frac{1}{\sqrt g} \frac{\delta S}{\delta V_\mu}\ .
\ee
These are components of the $R$-multiplet, which at the linear level have canonical couplings to the metric, to $A$ and to $V$, respectively. However, we remark that we do not set $A$ and $V$ to zero after having taken the variation, so these are currents in the presence of sources.

The expression in~\eqref{Esusy_bckgd_fields} shows that $E_{\rm susy}$ receives a contribution from the current in the $R$-multiplet in addition to the one from the temporal component of the energy-momentum tensor.
This is an alternate way to see that $E_{\rm susy}$ is different from the ordinary Casimir energy $E_0$ defined in eq.~\eqref{cas}. 

We now construct the charge associated with the time translation symmetry and compare it with~\eqref{Esusy_bckgd_fields}.
Recall that in the presence of background fields other than the metric, the energy-momentum tensor $T_{\mu\nu}$ is in general not conserved, $\nabla^\mu T_{\mu\nu}\neq 0$. This can be easily seen by revisiting the standard conservation proof (see {\it e.g.}\ Appendix E of~\cite{Wald:1984rg}) allowing for fields that do not satisfy their Euler-Lagrange equation. In particular, this applies to a supersymmetric field theory defined via the rigid limit of new minimal supergravity. One can see that the non-conservation equation of the energy-momentum tensor reads
\be
\nabla^\mu T_{\mu\nu} \ =\ (\diff A)_{\nu\mu}J_{\rm R}^\mu  -  \frac{3}{2}(\diff V)_{\nu\mu}J_{\rm FZ}^\mu   +  \frac{3}{2}  V_\nu \nabla_\mu J_{\rm FZ}^\mu\ ,
\ee
where we used that the $R$-current is conserved, $\nabla_\mu J_{\rm R}^\mu = 0$, while generically $\nabla_\mu J_{\rm FZ}^\mu \neq 0$.

Although $T_{\mu\nu}$ is not conserved, when the background admits a Killing symmetry generated by a vector $\xi^\mu$ one can introduce a modified energy-momentum current
\be
\timp_\xi^\mu\ =\ \xi^\nu  \Big(T_\nu{}^\mu + J_{\rm R}^\mu A_\nu - \frac{3}{2} J_{\rm FZ}^\mu V_\nu \Big)\ ,
\ee  
that {\it is} conserved, $\nabla_\mu \timp_\xi^\mu =0$. This is easily seen using ${\cal L}_\xi g = {\cal L}_\xi A = {\cal L}_\xi V = 0$. 
One can show that $\timp_\xi^\mu$ is just the canonical Noether current associated to the symmetry generated by $\xi$. Thus a conserved charge can be defined in the usual way.

We recall that two such Killing vectors exist in Euclidean backgrounds preserving two supercharges of opposite $R$-charge~\cite{Klare:2012gn,Dumitrescu:2012ha}. 
In Lorentzian signature, at least one null Killing vector exists~\cite{Cassani:2012ri}.
Our $M_{3}\times S^1_\beta$ background admits a Killing vector generating (Euclidean) time translations, $\xi = \frac{\partial}{\partial\tau}$. The associated conserved charge, to be identified with the Hamiltonian, is
\bea
 H &=& \int_{M_3} \diff^3 x \sqrt{g_{(3)}} \, u_\mu \timp^\mu_{\xi \,=\, \frac{\partial}{\partial\tau}} \nn \\ [2mm]
 &=&  \frac{1}{\beta}\int_{M_4}\diff^4 x \sqrt{g_{(4)}} \,\left( T^{\tau}{}_{\tau} + J_{\rm R}^\tau A_{\tau} - \frac{3}{2} J_{\rm FZ}^\tau V_{\tau}  \right)  \ ,\label{HamNewMin}
\eea
where $u^\mu$ is the unit time-like vector, that in the second line we expressed as $u = u_\mu \diff x^\mu = \sqrt{g_{\tau\tau}}\diff \tau$ (in the second line we also multiplied the three-dimensional integral by $1 = \frac{1}{\beta}\int_{S^1}\diff \tau$; since the charge is constant, this can be rewritten as a four-dimensional integral). 
The VEV of the final expression in~\eqref{HamNewMin} for $\beta\to \infty$ is immediately recognized as the $E_{\rm susy}$ given in~\eqref{Esusy_bckgd_fields}, that is what we wanted to show.

Finally, note that the Hamiltonian in (\ref{HamNewMin}) coincides with the supersymmetric $H_{\rm susy}$ used in section 2 only for the special choice $A_\tau =i/r_3$, which guarantees time-independent supersymmetry.


\section{Regularization of one-loop determinant}
\label{app:1LoopDet}

In this Appendix  we reconsider the regularization of the one-loop determinant for a free chiral multiplet on the Hopf surface $M^3 \times S^1$ (with $M_3 \simeq S^3$) computed in \cite{Assel:2014paa}.
It was found in \cite{DiPietro:2014bca} that the partition function on the Hopf surface has a universal behaviour in the small $\beta$ limit, where $\beta$ is the length of $S^1$.  In the case of the round $S^3 \times S^1$, this 
takes the form\footnote{Subleading terms in the small $\beta$ expansion were worked out in \cite{Ardehali:2015hya}.}
\be
\log Z^{\rm susy}_{S^3\times S^1_{\beta}} \ =\ \frac{16 \pi^2 r_3}{3 \beta} (c-a) + \scO(\beta^0) \, ,
\ee
where $r_3$ is the radius of $S^3$.
The result found in \cite{Assel:2014paa} are incompatible with this expansion, as in \cite{Assel:2014paa} the order ${\cal O}(\beta^{-1})$ vanishes instead. We propose here an alternative method to regularize one-loop determinant which on one hand agrees with the small beta expansion of \cite{DiPietro:2014bca} and on the other hand reproduces the large $\beta$ behaviour leading to the Casimir energy \eqref{squashedC}. Moreover, we show that this regularization method is equivalent to a cut-off regularization which manifestly preserves supersymmetry. 
The main difference between the procedure discussed  below and that  in \cite{Assel:2014paa} is that here all the  Kaluza-Klein (KK) modes on $S^1$ are dealt with in a manifestly symmetric way, while in \cite{Assel:2014paa} these were somewhat artificially split and combined with the Fourier modes on the three-sphere, in the triple gamma functions.  However, the precise reason for the discrepancy of the two  methods remains unclear.  While we were writing up 
this paper, \cite{Ardehali:2015hya} appeared, presenting a similar regularization method of one-loop determinants, leading to the result \eqref{squashedC} for the general deformed sphere.

One-loop determinant regularizations are based on the use of multiple gamma functions \cite{zbMATH01579914} and generalized zeta functions. The main mathematical tools that we will use are presented in a convenient way in appendices A and  B of \cite{Spiridonov:2010em}. We will also rely on \cite{FelderVarchenko}. 

The one-loop determinant suffers from UV divergences. It is given for a free chiral multiplet\footnote{Adding a gauge interaction with a flat gauge field as in the localization computation of \cite{Assel:2014paa} is straightforward.} by the formal expression
\begin{align}
Z^{\rm chiral}_{\rm 1{\textrm-}loop} &\equiv  Z =  \prod_{n_0 \in \bZ}  \ \prod_{n_1,n_2 \ge 0} 
\frac{ - \frac{r}{2} (a_1 +a_2) -  n_0  + (n_1+1) a_1 +  (n_2+1) a_2 }{    \frac{r}{2} (a_1 +a_2) +  n_0  + n_1 a_1 +  n_2 a_2}  \ , 
\label{1loopDet}
\end{align} 
where $a_1 = i b_1$, $a_2=i b_2$ in the notations of \cite{Assel:2014paa} and here we consider $b_1 > 0 , b_2> 0$. $r$ is the $R$-charge of the chiral multiplet. The analysis of this paper allows to re-interpret each factor appearing in the numerator of \eqref{1loopDet} as the contribution from a single Fermi multiplet to the one-loop determinant and each factor in the denominator as the contribution of a single chiral multiplet. A crucial difference with the Hamiltonian quantization analysis is that now we have a tower of KK modes on $S^1$ with KK level parameterized by $n_0 \in \bZ$.

\subsection{Two-step regularization}

Here we proceed by first regularizing the sum over $n_1,n_2 \ge 0$ with double gamma functions for a fixed $n_0$, which corresponds to the one-loop determinant on $M_3$ of the $n_0$-th  KK 
mode along $S^1$, and then regularizing the infinite sum over $n_0 \in \bZ$. We start with
\begin{align}
Z &= \prod_{n_0 \in \bZ}  F [ u + n_0 ]  \nn\\
F[v] &= \prod_{n_1,n_2 \ge 0} 
\frac{ -v  + (n_1+1) a_1 +  (n_2+1) a_2 }
{  v  + n_1 a_1 +  n_2 a_2}  \ = \  \frac{\Gamma_2(v,a_1,a_2)}{\Gamma_2(a_1+a_2 - v ,a_1,a_2)}  \nn\\
&=  \Gamma_h(v,a_1,a_2)  \ , 
\label{Zu}
\end{align}
where $ u  =   \frac{r}{2} (a_1 +a_2)$ and the hyperbolic gamma function is defined by
\begin{align}
&\Gamma_h(v,a_1,a_2)  =   {e}^{\frac{\pi i}{2} B(v,a_1,a_2)} \, \frac{({e}^{2\pi i (a_1 - v)/a_2} ; {e}^{2\pi i a_1/a_2})_{\infty} }{({e}^{-2\pi i  v/a_1} ; {e}^{-2\pi i a_2/a_1})_{\infty}} 
\label{Gh1}\\
& (x,q)_{\infty}  =   \prod_{k \ge 0} (1- x q^k) \, , \qquad  B(v,a_1,a_2)  =   \frac{1}{a_1 a_2} \lp  \lp v - \tfrac{a_1+a_2}{2}\rp^2  - \tfrac{a_1^2 + a_2^2}{12} \rp \, . \nn
\end{align}
The hyperbolic gamma function admits also the following representation \cite{Spiridonov:2010em}
\begin{align}
&\Gamma_h(v,a_1,a_2) \ = \  {e}^{-\frac{\pi i}{2} B(v,a_1,a_2)} \, \frac{({e}^{2\pi i (v- a_2)/a_1} ; {e}^{-2\pi i a_2/a_1})_{\infty} }{({e}^{2\pi i  v/a_2} ; {e}^{2\pi i a_1/a_2})_{\infty}}  \, .
\label{Gh2}
\end{align}
Using \eqref{Gh1} for $n_0 \ge 1$ and \eqref{Gh2} for $n_0 \le 0$, we obtain
\begin{align}
Z \ & =\ {e}^{\frac{\pi i}{2} C(u,a_1,a_2)} \, \frac{\Gamma_e  \lp \frac{u}{a_2}, - \frac{1}{a_2}, \frac{a_1}{a_2}  \rp }{\Gamma_e  \lp \frac{u-a_2}{a_1}, - \frac{1}{a_1}, -\frac{a_2}{a_1}  \rp}  \ = \ {e}^{\frac{\pi i}{2} C(u,a_1,a_2)} \, {e}^{\frac{\pi i}{2} D(u,a_1,a_2)} \,  \Gamma_e(u,a_1,a_2) \ ,
\label{ZGammaE}
\end{align}
where the elliptic gamma function $\Gamma_e$  is defined by
\begin{align}
\Gamma_e(x,\tau,\sigma)= \prod_{n_1,n_2 \ge 0} \frac{1- {e}^{2\pi i (-x + (n_1+1) \tau + (n_2+1) \sigma)} }{1- {e}^{2\pi i (x + n_1 \tau + n_2 \sigma)}} \ .
\label{defGammae}
\end{align}
The last equality in \eqref{ZGammaE} is obtained from the modular properties of the elliptic gamma function \cite{Spiridonov:2010em,FelderVarchenko} and the functions $C,D$ are given by
\begin{align}
D(u,a_1,a_2) &= \frac{1}{a_1a_2} \lp   \frac{2}{3} (u')^3 + (u')^2 + \frac{2- a_1^2 -a_2^2}{6} \, u'   - \frac{a_1^2 + a_2^2}{12}\rp\,,
 \\
C(u,a_1,a_2) &=  \frac{1}{a_1a_2} \lp \sum_{n\ge 1} \Big[  (u'+n)^2   - \frac{a_1^2 + a_2^2}{12} \Big]  
-  \sum_{n\ge 0} \Big[  (- u'+n)^2   - \frac{a_1^2 + a_2^2}{12} \Big] \rp  \, ,
\label{Cinfty}
\end{align}
with $u' = u - \frac{a_1 +a_2}{2} =   \frac{r-1}{2} (a_1 +a_2)$. Note that $D (u,a_1,a_2)$ is a well defined function, but $C(u,a_1,a_2)$
involves an infinite sum over $n$, which needs to be regularized.  We do this
 using the standard Riemann zeta function,\footnote{This is similar to the regularization presented in appendix B of \cite{Aharony:2013dha}.} which is compatible with the partial cancellation of terms between the two infinite sums:
\begin{align}
C(u,a_1,a_2) &=  \frac{1}{a_1a_2} \lp  - (u')^2  + \frac{a_1^2 + a_2^2}{12} + 4 u'  \sum_{n\ge 1} n  \rp  
 = \frac{1}{a_1a_2} \lp  - (u')^2  + \frac{a_1^2 + a_2^2}{12} - \frac{u'}{3} \rp \ ,
\label{Creg}
\end{align} 
where we used  $\sum_{n\ge 1} n = \zeta(-1) = -\frac{1}{12}$. Notice that despite the formal similarity of the infinite sums in (\ref{Cinfty}) and (\ref{CasimirRound}), the regularization of the two sums is performed using two \emph{different} prescriptions. 
The final result is
\begin{align}
& Z   =   {e}^{i \pi \Psi(u,a_1,a_2) } \, \Gamma_e(u,a_1,a_2)   \, , \qquad  \Psi(u,a_1,a_2)  =  \frac{(u')^{3}}{3 a_1a_2} - \frac{ a_1^2 + a_2^2}{12 a_1a_2} \, u' \, .
 \label{Regul2}
\end{align}
This reproduces the supersymmetric Casimir energy \eqref{squashedC}  and it can be shown that is compatible with the small $\beta$ limit of \cite{DiPietro:2014bca}, which is obtained by setting $a_1 = i \frac{\beta}{2\pi r_3} \fb$, $a_2= i \frac{\beta}{2\pi r_3}  \fb^{-1}$ and keeping $\fb$ fixed in the limit. 
The difference with the result of \cite{Assel:2014paa} is only the absence of the term $\frac{u'}{6 a_1 a_2} =  \frac{r-1}{12} \frac{a_1 +a_2}{a_1a_2}$ in $\Psi(u,a_1,a_2)$, which affects the small $\beta$ limit, but not the large $\beta$ limit.  

\subsection{Cut-off regularization}

We present now a different  method of regularization which uses a cut-off on the momentum modes. The treatment of the KK modes in this method is more transparent and the result agrees with  \eqref{Regul2}, thus corroborating it.

A reliable way of regularizing the one-loop determinant is to introduce a cut-off on the momentum of the modes or more generally a smooth truncation of the modes above a momentum scale $\Lambda$ (this means that the contributions of modes above $\Lambda$ are counted with decreasing  weights). Note that this preserves supersymmetry since each momentum in the product carries the contribution of a specific multiplet in supersymmetric quantum mechanics. The (logarithm of the) regularized determinant can be expanded in powers of the cut-off at large $\Lambda$. The diverging contributions are expected to be removed by local supersymmetric counterterms, while the finite piece yields the regularized result, possibly subject to ambiguities. In the case at hand we compute the partition function on a curved manifold admitting two Killing spinors of new minimal supergravity of opposite chirality and it was shown in \cite{Assel:2014tba} that there is no ambiguity in the finite result.

For simplicity, we focus here  on the case of the round $S^3 \times S^1$, which amounts to setting
 $a_1=a_2=a \in i \bR_{>0}$.
In this case, the chiral multiplet one-loop determinant simplifies to
\begin{align}
Z &= \prod_{n_0 \in \bZ}  \ \prod_{n_1,n_2 \ge 0} 
\frac{ -u -  n_0  + (n_1+ n_2 +2) a }
{  u +  n_0  + (n_1 + n_2) a} \ = \ \prod_{n_0 \in \bZ}  \ \prod_{m \ge 0} 
\lp \frac{ -u -  n_0  + (m +2) a }
{  u +  n_0  + m a} \rp^{m+1}  \nn\\
&= \prod_{n_0 \in \bZ} F[u+n_0] \ ,
\end{align} 
with $u = ra$ and
\begin{align}
\log F[v] &= \sum_{m \ge 0} (m+1) \big[ \log(-v+(m+2)a) - \log(v+m a) \big] \ .
\end{align}
We regularize the sum by introducing a cut-off $\frac{\beta}{2\pi} \Lambda$ on the $S^1_{\beta}$ KK momentum level $n_0$ and $\frac{\beta}{2\pi |a|} \Lambda$ on the $S^3$ KK momentum level $m$.\footnote{The relative factor $|a| = \frac{\beta}{2\pi r_3}$ takes into account the ratio of scales between $S^1_{\beta}$ and $S^3$ KK modes.}  To simplify slightly the notations we simply set $\beta= 2\pi$ in the following.
The regularization is done by using any smooth decreasing function $f$ such that $f(0)=1$ and going to zero at infinity sufficiently fast to make the sum converge. The regularized sum over $S^3$ modes is
\begin{align}
\log F[v] &= \sum_{m \ge 0} (m+1) \big[ \log(-v+(m+2)a) - \log(v+m a) \big]  f \lp \frac{m}{\Lambda |a^{-1}|} \rp  \ .
\end{align}
We make use of the Euler-MacLaurin formula\footnote{The Euler-MacLaurin formula for a convergent sum reads:
\begin{align}
\sum_{n \ge 0} g(n) &=  \int_0^{\infty} g(x) \diff x   +  \frac 12 g(0)   +  \sum_{k \geq 1} \frac{\zeta(1-2k)}{(2k-1)!} \, g^{(2k-1)}(0) \, .
\label{EulerMacLaurin}
\end{align}} 
to work out the large $\Lambda$ expansion
\begin{align}
\log F[v] \ &=\ 2 \Lambda |a^{-1}| (1 - v a^{-1}) \int_0^{\infty} f + \scO(\Lambda^0) \nn \\
\ &=\ c_1 \Lambda a^{-2} (v-a) + \scO(\Lambda^0) \ ,
\end{align}
with $c_1 =  2 i \int_0^{\infty} f$.
The finite part of the large $\Lambda$ expansion is complicated as it receives an infinite number of contributions from the Euler-MacLaurin expansion, however it is easy to see that it is independent of the function $f$ and we can trust that it will reproduce the (logarithm of the) hyperbolic gamma function at $a_1=a_2$, since this is the well-known result for the $3d$ chiral multiplet one-loop determinant. We obtain
\be
 F[u + n_0] \ \simeq\ {e}^{ c_1 \Lambda a^{-2} (u- a +n_0) } \, \Gamma_h(u+n_0,a,a) \ ,
\ee
where $\simeq$ indicates that we dropped the term $\scO(\Lambda^{-1})$ in $\log  F[u + n_0]$.
The same manipulations as in the regularization above yield the result
\begin{align}
Z\ & =\ {e}^{\frac{\pi i}{2} \ti C(u,a,a)} \, {e}^{\frac{\pi i}{2} D(u,a,a)} \,  \Gamma_e(u,a,a)  \ ,
\end{align}
with 
\begin{align}
\ti C(u,a,a) \ &= \  \frac{1}{a^2} \sum_{n\ge 1} \Big[  (u'+n)^2  - \frac{a^2}{6} + c_1 \Lambda (u'+n)  \Big]\, \hat f \lp \frac{n}{\Lambda} \rp \nn\\
& \quad 
+  \frac{1}{a^2} \sum_{n\ge 0} \Big[  - (u'-n)^2   + \frac{a^2}{6}  + c_1 \Lambda (u'-n) \Big]\, \hat f \lp \frac{n}{\Lambda} \rp   \, ,
\label{tiC}
\end{align}
where  $u' = u - a = (r-1)a$ and $\hat f$ is a second regulating function implementing the smooth cut-off on $S^1$ KK modes. 
Applying again the Euler-MacLaurin formula we obtain
\begin{align}
\ti C(u,a,a) \ &=\ \frac{1}{a^2} \lp  c_2 \, \Lambda^2   u' -  u'{}^2 - \frac{u'}{3} + \frac{a^2}{6} \rp~,
\end{align}
with $c_2 = 4 \int_0^{\infty} y \hat f(y) \diff y + \frac{8 c_1}{i\pi} \int_0^{\infty} \hat f(y) \diff y$. The finite piece reproduces the result \eqref{Creg} for $a_1=a_2=a$, so that the cut-off regularization result $Z_{\rm reg}$ will match the regularization described above. 
The diverging piece should be removed with a dimension two supersymmetric counterterm constructed with new minimal supergravity background fields \cite{Assel:2014tba}.
We have obtained
\begin{align}
\log Z \ &\simeq  \ \frac{i \pi}{2} c_2 \, \Lambda^2 \, (r-1) a^{-1}  + \log Z_{\rm reg} \ .
\label{Ztot}
\end{align}
There is a single supergravity term of mass dimension two in new minimal supergravity~\cite{Assel:2014tba}; it is the usual Einstein-Hilbert new minimal supergravity action, that we can take with coefficient $\Lambda^2$. Its bosonic part is given by
\be\label{LagrangianNewMin}
S^{(4)}_{R} \ = \ \frac{\Lambda^2}{2} \int \diff^4 x \, \sqrt g \left(R + 6 V_\mu V^\mu - 8 A_\mu V^\mu \right)\, .
\ee
Its evaluation on the $S^3 \times S^1$ background \cite{Dumitrescu:2012ha} yields\footnote{Note that the result is in units where the $S^1$ radius is fixed to one. To reinstate the $\beta$ dependence, one can shift $\Lambda \rightarrow \frac{\beta}{2\pi}\Lambda$.}
\begin{align}
 S^{(4)}_{R} \ &\propto  \ \Lambda^2  \, a^{-2}  \kappa \ ,
\end{align}
where the constant $\kappa$ parametrizes a freedom in the choice of background. The choice $\kappa = a$ is the most natural,  because it preserves the $SU(2)_l \times SU(2)_r$ isometries of $S^3$. 
This is precisely the term needed to remove the divergent piece in \eqref{Ztot}. We conclude that the cut-off regularization further validates the regularization method presented above.


\bibliographystyle{JHEP}
\bibliography{Newbib}

\end{document}